\documentclass[trackchanges]{aastex7}

\usepackage{graphicx}
\usepackage{booktabs}
\usepackage{tabularx}
\usepackage{graphicx}
\usepackage{subcaption}
\usepackage{dcolumn}
\usepackage{bm}

\usepackage{rotating}
\usepackage{subcaption}
\usepackage{float}
\usepackage{amsmath}
\usepackage{hyperref}
\hypersetup{
    colorlinks=true,
    linkcolor=blue,
    filecolor=magenta,      
    urlcolor=red,
    pdftitle={Overleaf Example},
    pdfpagemode=FullScreen,
    }
\usepackage{color}
\hypersetup{
    colorlinks=true,
    linkcolor=red,
    citecolor=blue,
}

\begin{document}


\title{$E_{\rm peak}$– $\alpha$ Correlation in Time Resolved GRB Spectra: A Bottom-Up Approach with Optically Thin Inverse Compton Scattering Model}

\author[0009-0005-6858-6356]{Pragyan Pratim Bordoloi} 
\affiliation{School of Physics, Indian Institute of Science Education and Research Thiruvananthapuram, Thiruvananthapuram, 695551, India}
\email[show]{pragyan.bordoloi21@iisertvm.ac.in}

\author[0009-0008-2096-5797]{Ayush Shivkumar}
\affiliation{School of Physics, Indian Institute of Science Education and Research Thiruvananthapuram, Thiruvananthapuram, 695551, India}
\email[show]{}  

\author[0000-0003-3220-7543]{Shabnam Iyyani}
\affiliation{School of Physics, Indian Institute of Science Education and Research Thiruvananthapuram, Thiruvananthapuram, 695551, India}
\affiliation{Centre for High Performance Computing, Indian Institute of Science Education and Research Thiruvananthapuram, Thiruvananthapuram, 695551, India}
\email[show]{shabnam@iisertvm.ac.in}

\begin{abstract}
Gamma-ray bursts (GRBs) are the brightest explosions in the Universe, yet the origin of their emission remains uncertain. Time-resolved spectral analysis offers key insights into the evolution of spectral shapes, constraining both radiation mechanisms and emission-site microphysics. Observationally, GRB spectra are well described by the empirical Band function,characterized by the peak energy ($E_{\mathrm{peak}}$) and low-energy spectral index ($\alpha$). We investigate the temporal evolution of spectra produced by optically thin inverse-Compton scattering (ICS) within a standard fireball jet framework, focusing on the scenarios that can produce the two commonly observed spectral evolution patterns: hard-to-soft evolution and intensity tracking, within a single emission pulse. The evolution is analysed using both Bayesian block and constant-fluence binning, with the observed spectrum modeled consistently using the Band function. Using this bottom-up approach, we find that optically thin ICS yields a positive $E_{\mathrm{peak}}$–$\alpha$ correlation, with $\alpha$ evolving from hard (Planck-like, $> +0.5$) to softer ($< -0.67$) values. Such hard $\alpha$ values are inconsistent with standard synchrotron emission. This characteristic evolution in the $E_{\mathrm{peak}}$–$\alpha$ plane, therefore, provides a diagnostic signature of optically thin ICS as the dominant radiation mechanism during the prompt phase of GRBs. Furthermore, this type of smooth evolution of $\alpha$ within a single pulse does not require invoking a transition between different radiation mechanisms, unless additional observational evidence supports such a change.


\end{abstract}



\section{Introduction} 
GRBs are extremely bright, short-lived flashes of gamma rays lasting from milliseconds to several hundred seconds. During the prompt emission phase, GRBs exhibit rapid and pronounced spectral evolution, particularly within individual pulses. Time-resolved 
spectral analysis of GRB prompt emission is commonly performed using empirical models such as the Band function \citep{Band_1993}, which effectively characterizes both the spectral shape and its temporal evolution. The Band function consists of two smoothly connected 
power laws with a characteristic peak energy, $E_{\mathrm{peak}}$, while $\alpha$ and $\beta$ denote the low- and high-energy spectral indices, respectively. Among these, $\alpha$ and $E_{\mathrm{peak}}$ are generally the best constrained parameters and serve as key tracers of spectral evolution.

The low-energy spectral index, $\alpha$, is closely linked to the underlying radiation mechanism, whereas $E_{\mathrm{peak}}$ represents the characteristic energy scale of the emission process. Low-energy slopes with $\alpha \lesssim -0.67$ are broadly consistent 
with optically thin synchrotron emission \citep{Preece1998,Sari1998}, while harder spectra challenge the simplest synchrotron models and often motivate alternative scenarios such as photospheric emission, Comptonization, or modified synchrotron models. In synchrotron scenarios, $E_{\mathrm{peak}}$ is associated with the characteristic synchrotron frequency 
produced by electrons near the characteristic Lorentz factor $\gamma_m$. In inverse Compton or Comptonization models, $E_{\mathrm{peak}}$ reflects the characteristic energy of photons upscattered by the electron population and depends on the properties of both the seed photon field and the scattering electron distribution.

Under steady-state radiative conditions within each time bin, no intrinsic correlation between $\alpha$ and $E_{\mathrm{peak}}$ is generally expected in time-resolved spectral evolution studies. However, observations have revealed correlations between these 
parameters within individual pulses. Studies such as \citep{Kaneko2006} show that nearly $26\%$ of BATSE GRBs exhibit a positive $\alpha$--$E_{\mathrm{peak}}$ correlation within individual pulses, with both parameters often tracking the flux evolution. In contrast, 
population-wide analyses, such as the \textit{Fermi} GBM spectral catalog \citep{Gruber_2014}, report no significant global correlation across the GRB population. 
This disparity suggests that the observed $\alpha$--$E_{\mathrm{peak}}$ relation is primarily driven by pulse-level physical evolution rather than intrinsic burst-to-burst differences.

In reality, GRB prompt emission is highly dynamic, as reflected in the irregular and rapidly varying light curves. Physical conditions within the emission region, including the 
thermal flux or photospheric temperature, electron acceleration efficiency, magnetic field strength, and optical depth, can evolve significantly on short timescales. These variations are likely governed by changes in jet dynamics, such as fluctuations in injection 
luminosity, central engine activity, baryon loading, and the dissipation process itself. Consequently, even within a single radiation mechanism, whether synchrotron emission, inverse Compton scattering, or subphotospheric Comptonization, temporal evolution may 
imprint characteristic and potentially distinguishable signatures on the evolution of $\alpha$ and $E_{\mathrm{peak}}$ throughout an individual pulse.

This motivates the central objective of the present work to investigate whether different radiation mechanisms naturally produce distinct $\alpha$-$E_{\mathrm{peak}}$ correlations 
when realistic temporal evolution is incorporated, and whether such correlations can serve as diagnostic probes of the underlying emission process. In this paper, we focus specifically on optically thin inverse Compton scattering (ICS) as a candidate radiation mechanism for GRB prompt emission. We employ a bottom-up simulation framework in which 
synthetic time-resolved spectra are generated using physically motivated electron distributions within the baryonic fireball framework \citep{Meszaros2006, Kumar_Zhang2015, Iyyani_2018}, with radiative processes implemented using the Naima \citep{naima} and 
ThreeML \citep{3ml} packages. The simulated spectra are subsequently fitted with the Band function, closely replicating standard observational analysis procedures and enabling the extraction of the corresponding $\alpha$ and $E_{\mathrm{peak}}$ values. 
The overarching goal of this work is to investigate whether the $E_{peak} - \alpha$ evolution, including its characteristic range and systematic evolution of $\alpha$ with $E_{peak}$, can be naturally 
produced within the framework of optically thin ICS, and to assess whether this joint spectral behaviour can serve as a useful diagnostic for evaluating ICS as a viable radiation mechanism for GRB prompt emission.

The paper is organized as follows. In Section~(\ref{Simul_sec}), we describe the simulation framework for time-resolved optically thin inverse Compton spectra and discuss the explored physical parameter space. In Section~(\ref{Time_resol_spec_analy}), we present the time-
resolved spectral analysis of the simulated spectra using different binning methods. In Section~(\ref{results}), we present the results on the $\alpha$--$E_{\mathrm{peak}}$ correlation, followed by a discussion in Section~(\ref{Discussion}). Finally, in Section~(\ref{conclusion}), we summarize our findings and discuss their broader implications.

\section{Simulation of optically thin inverse Compton scattering radiation}
\label{Simul_sec}
\subsection{Input Parameters For Simulation}
\label{Input_para}
In this study, we consider inverse Compton scattering between thermal photons originating from the photosphere and a hybrid electron population located at an optically thin dissipation region within the GRB outflow. The electron distribution consists of a thermal component together with a non-thermal power-law tail. We adopt the same hybrid electron distribution described in \citealt{Bordoloi_Iyyani_2025}, characterized by three independent parameters: $\theta$, $\delta$, and $n_0$, where $\theta$ is the dimensionless electron temperature, $\delta$ is the power-law index of the accelerated electron component, and $n_0$ is the normalization of the electron distribution. For convenience, we reparameterize the electron temperature in terms of the characteristic thermal energy scale,
\begin{equation}
    E_{\rm th} = \theta m_e c^2,
\end{equation}
where $E_{\rm th}$ represents the characteristic thermal energy of the electrons.

The seed photon field is assumed to be a pure blackbody spectrum characterized by the temperature $kT_{\rm ph}^{\rm co}$. These thermal photons originate at the photosphere ($R_{ph}$) and and propagate outward to the dissipation site ($R_{d}$), where they undergo inverse Compton scattering with the electron population. The thermal photon flux decreases geometrically as $(R_{\rm ph}/R_d)^2$ due to spherical expansion. In addition, only a fraction $(1-e^{-\tau}) \simeq \tau$ of the photons undergo scattering at an optically thin dissipation site. For a relativistic outflow, the optical depth above the photospheric radius scales approximately as $\tau(R_d)\sim R_{ph}/R_d$. Consequently, the effective seed photon flux participating in the scattering process is reduced approximately by a factor $(R_{ph}/R_d)^3$.

The simulated inverse Compton spectra are generated in the co-moving frame and thereby they need to be transformed to the observer frame to evaluate the observed $E_{\mathrm{peak}}$ and $\alpha$. This transformation incorporates relativistic Doppler boosting due to the bulk motion of the outflow and cosmological redshift effects.

{\bf Transformation to the Observer Frame :}

The relation between co-moving (primed quantities) and observer-frame quantities is given by
\begin{equation}
    t_{\rm obs} = \frac{1+z}{\mathcal{D}}\,t', \qquad
    E_{\rm obs} = \frac{\mathcal{D}}{1+z}\,E'_{\rm IC},
    \label{eq:transformations}
\end{equation}
where
\begin{equation}
    \mathcal{D}
    =
    \frac{1}{\Gamma(1-\beta\cos\theta)}
\end{equation}
is the Doppler factor, $\beta$ is the dimensionless bulk velocity, $\theta$ is the angle between the outflow velocity and the observer line of sight, and $z$ is the redshift.

{\it Naima} computes the inverse Compton emissivity assuming isotropic emission in the co-moving frame. For an ultra-relativistic outflow viewed on-axis, relativistic aberration confines the emission within a cone of angular width $\sim 1/\Gamma$. Upon integration over the relativistically beamed solid angle, the Doppler factors associated with the energy and time transformations cancel in the photon number spectrum. Consequently, the photon number spectrum per unit time and energy remains invariant between the co-moving and observer frames, apart from the energy shift and cosmological redshift corrections \citep{Bordoloi_2026}, giving:
\begin{equation}
    \frac{{\rm d}N}{{\rm d}t_{\rm obs}\,{\rm d}E_{\rm obs}}
    =
    \frac{{\rm d}N'}{{\rm d}t'\,{\rm d}E'_{\rm IC}}.
\end{equation}

Within the baryonic fireball framework \citep{Pe'er2007,Iyyani2013}, the co-moving photospheric temperature $T^{'}_{\rm ph}$ is determined by the burst luminosity $L$, bulk Lorentz factor $\Gamma$, and nozzle radius $R_0$:
\begin{equation}
T^{'}_{\rm ph}(R_{ph}) 
= 
\left( \frac{L}{4\pi R_{0}^{2} c a} \right)^{1/4}
\Gamma^{-1}
\left( \frac{R_{ph}}{R_{s}} \right)^{-2/3},
\label{comoving_Trph}
\end{equation}
where $R_{ph}$ is the photospheric radius,
\begin{equation}
R_{ph} =
\frac{L\sigma_T}{8\pi m_p c^3 \Gamma^3},
\label{Photo_radius_1}
\end{equation}
$\sigma_T$ is the Thomson cross section, and the saturation radius is given by $R_s = \Gamma R_0$ considering the photosphere forms above $R_s$. Thus, specifying $L$, $\Gamma$, and $R_0$ uniquely determines the seed photon temperature $kT$.

The full parameter set required to simulate the inverse Compton spectra therefore consists of the electron distribution parameters $E_{\rm th}$, $\delta$, and $n_0$, together with the bulk Lorentz factor $\Gamma$, nozzle radius $R_0$, burst luminosity $L$, dissipation radius $R_d$, and luminosity distance $d_L$ (or equivalently the redshift $z$). Throughout this work, the electron power-law index is fixed at the canonical value $\delta = 2.23$.
\subsection{Time-Dependent Parameter Space}
\label{Physical_para_setup}
The model input parameters are intrinsically linked to the observed spectral properties of GRBs, particularly the total energy flux and the spectral peak energy, $E_{\rm peak}$. Consequently, for a time-evolving pulse, 
the temporal evolution of these observables constrains the physically allowed parameter space. The input parameters must therefore be chosen such that the 
resulting inverse Compton scattering spectra remain consistent with the observed spectral evolution. In the following, we discuss the adopted parameter evolution and the associated physical constraints.  \\

\begin{figure*}[!ht]
    \centering
    \includegraphics[width=0.7\linewidth]{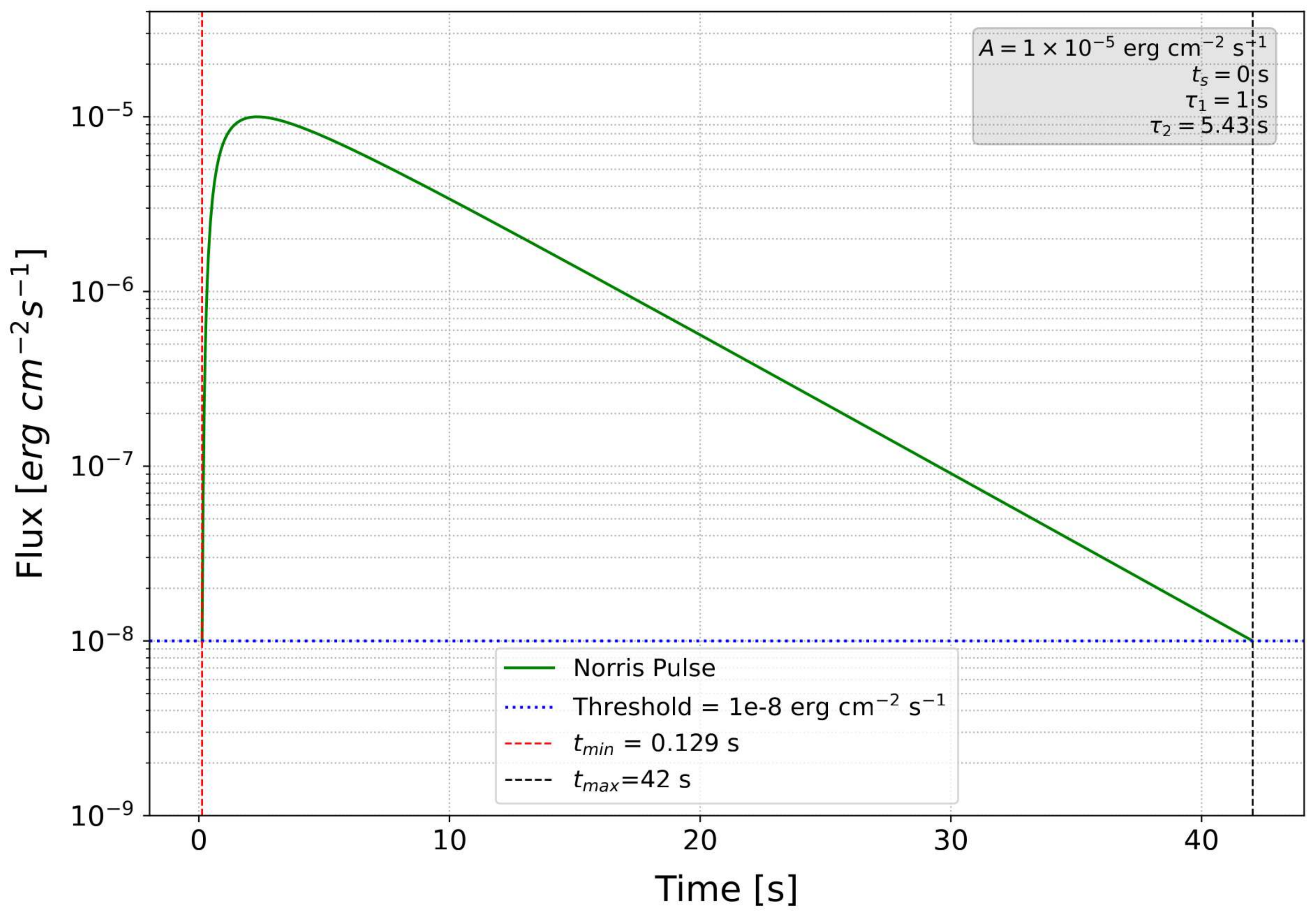} 
    \caption{The Norris pulse profile adopted for modeling the temporal evolution of the energy flux is shown. The vertical red dashed lines indicate $t_{\min}$ and $t_{\max}$, which define the observable time interval of the emission pulse above the \textit{Fermi} GBM detection threshold. 
    }
    
    \label{Norris_pulse}
\end{figure*}

{\bf (I) Pulse profiles for the energy integrated flux:}
\label{Norris_section}
GRB light curves typically comprise one or more emission pulses exhibiting a fast-rise, exponential-decay (FRED)-like morphology, reflecting the highly variable nature of the prompt emission. These individual pulses are well 
described by the empirical Norris function \citep{Norris_2005}, given by Equation~(\ref{Norris_fnc}), which provides an effective representation of GRB pulse profiles. Since the energy-integrated flux is also expected to track the pulse evolution, we adopt the Norris function to model its temporal behaviour as follows: 
\begin{equation}
F_E(t) = A \exp \left[
2 \sqrt{\frac{\tau_1}{\tau_2}}
- \left(
\frac{\tau_1}{t - t_s}
+ \frac{t - t_s}{\tau_2}
\right)
\right],
\label{Norris_fnc}
\end{equation}
where $F_E(t)$ is the observed energy flux as a function of time $t$. The parameter $A$ denotes the pulse amplitude with units of energy flux, and $t_s$ represents the start time of the pulse. The parameter $\tau_1$ characterizes the pulse rise time, controlling how rapidly the flux increases from the onset of the pulse, while $\tau_2$ corresponds to the decay time, governing the exponential decline of the pulse after the peak. \\
In our analysis, we adopt the following Norris function parameters: $\tau_1 = 1~\mathrm{s}$, $\tau_2 = 5.43~\mathrm{s}$, $t_s = 0~\mathrm{s}$, and an amplitude $A = 1 \times 10^{-5}~\mathrm{erg~cm^{-2}~s^{-1}}$. The Norris pulse corresponding to this set of parameters is shown in Figure~\ref{Norris_pulse}. The chosen value of the amplitude $A$ ensures a sufficiently bright pulse, 
providing an adequate number of photon counts for reliable statistical analysis. The characteristic timescales $\tau_1$ and $\tau_2$ are motivated by the values reported by \citealt{Norris_2005}, who applied the Norris pulse function to a sample of 24 GRBs comprising both single- and multi-pulsed light curves. The study found 
representative average timescales of $\tau_1 = 8.55~\mathrm{s}$ and $\tau_2 = 5.43~\mathrm{s}$. However, since the present analysis focuses specifically on fast-rise exponential-decay (FRED)-like pulses, we consider the 
rise timescale to be shorter than the decay timescale, i.e., $\tau_1 < \tau_2$. Accordingly, we adopt $\tau_1 = 1~\mathrm{s}$, which produces a sharply rising pulse profile consistent with the FRED morphology.

The pulse duration is defined as the time interval between $t_{\min}$ and $t_{\max}$, corresponding to the epochs before and after the pulse peak at which the flux falls below the \textit{Fermi} GBM detection threshold 
of $1 \times 10^{-8}~\mathrm{erg\,cm^{-2}\,s^{-1}}$ \citep{Meegan2009}. For the adopted Norris pulse parameters, this yields a pulse duration of approximately $42~\mathrm{s}$, as shown in Figure~\ref{Norris_pulse}.

{\bf (II) Types of $E_{ peak}$ temporal evolution patterns:}
\label{Epeak_evol}
Observational studies have shown that the spectral peak energy of GRBs, characterised by the Band-function parameter $E_{\rm peak}$, exhibits two primary types of 
temporal evolution: (i) a hard-to-soft evolution, in which $E_{\rm peak}$ decreases monotonically with time over the 
burst duration, largely independent of the flux evolution, and (ii) an intensity-tracking behaviour, where $E_{\rm peak}$ closely follows the burst intensity (or flux) \citep{Lu2010}.
\\

\begin{figure}[htbp]
\centering
\begin{minipage}{0.47\textwidth}
\setlength{\unitlength}{1cm}
\begin{picture}(0,0)
    \put(-0.36,6){\small\textbf{(a)}}
\end{picture}
\includegraphics[width=\linewidth]{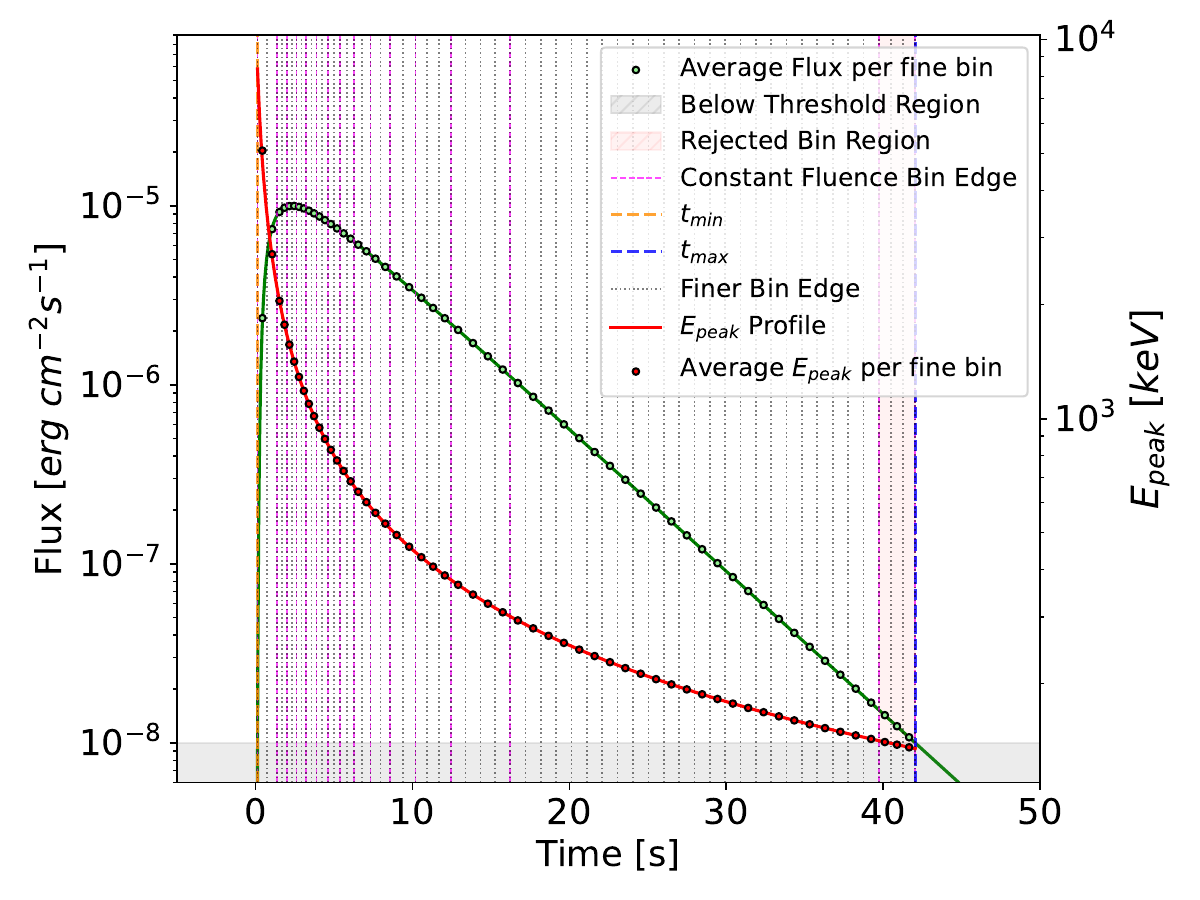}
\end{minipage}
\hspace{0.01\textwidth}
\begin{minipage}{0.47\textwidth}
\setlength{\unitlength}{1cm}
\begin{picture}(0,0)
    \put(-0.38,6){\small\textbf{(b)}}
\end{picture}
\includegraphics[width=\linewidth]{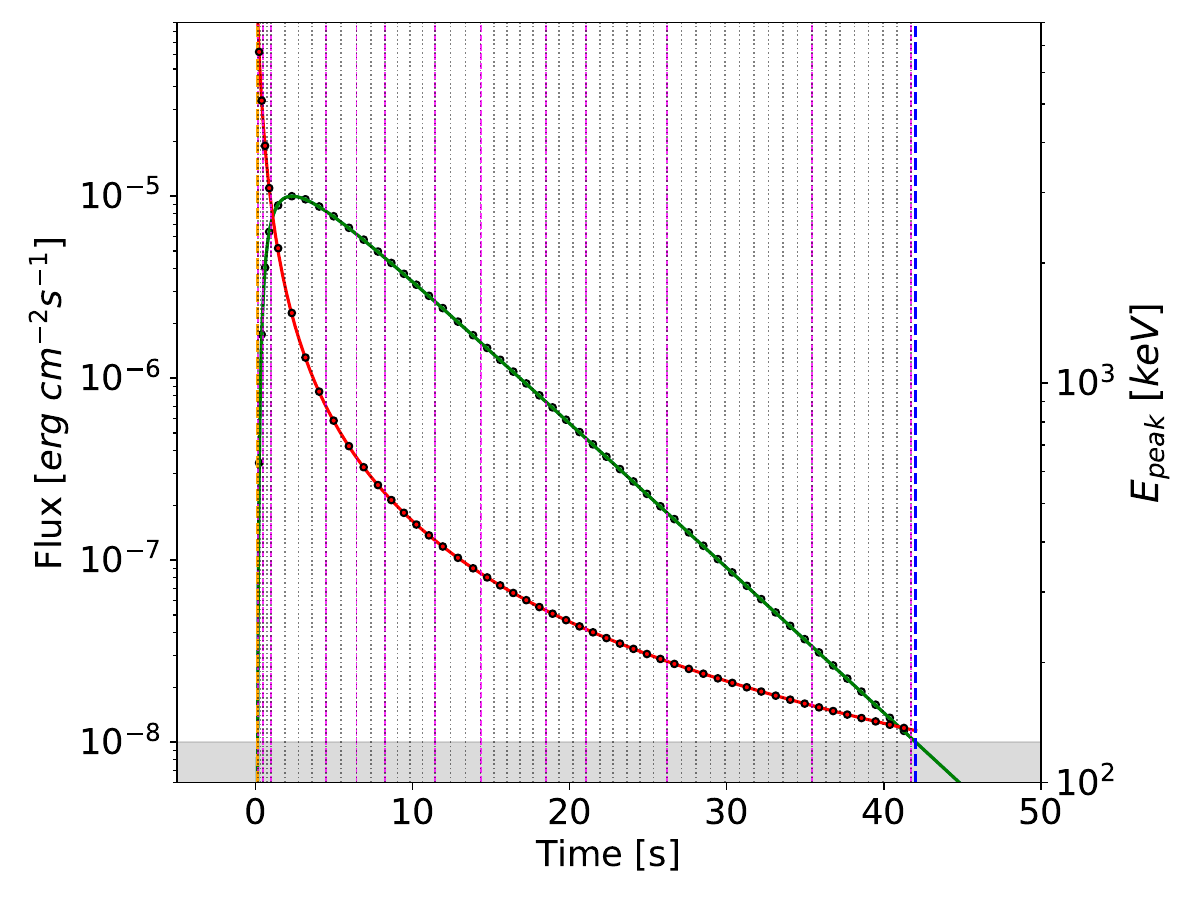}
\end{minipage}

\vspace{0.3cm}

\begin{minipage}{0.47\textwidth}
\setlength{\unitlength}{1cm}
\begin{picture}(0,0)
    \put(-0.36,6){\small\textbf{(c)}}
\end{picture}
\includegraphics[width=\linewidth]{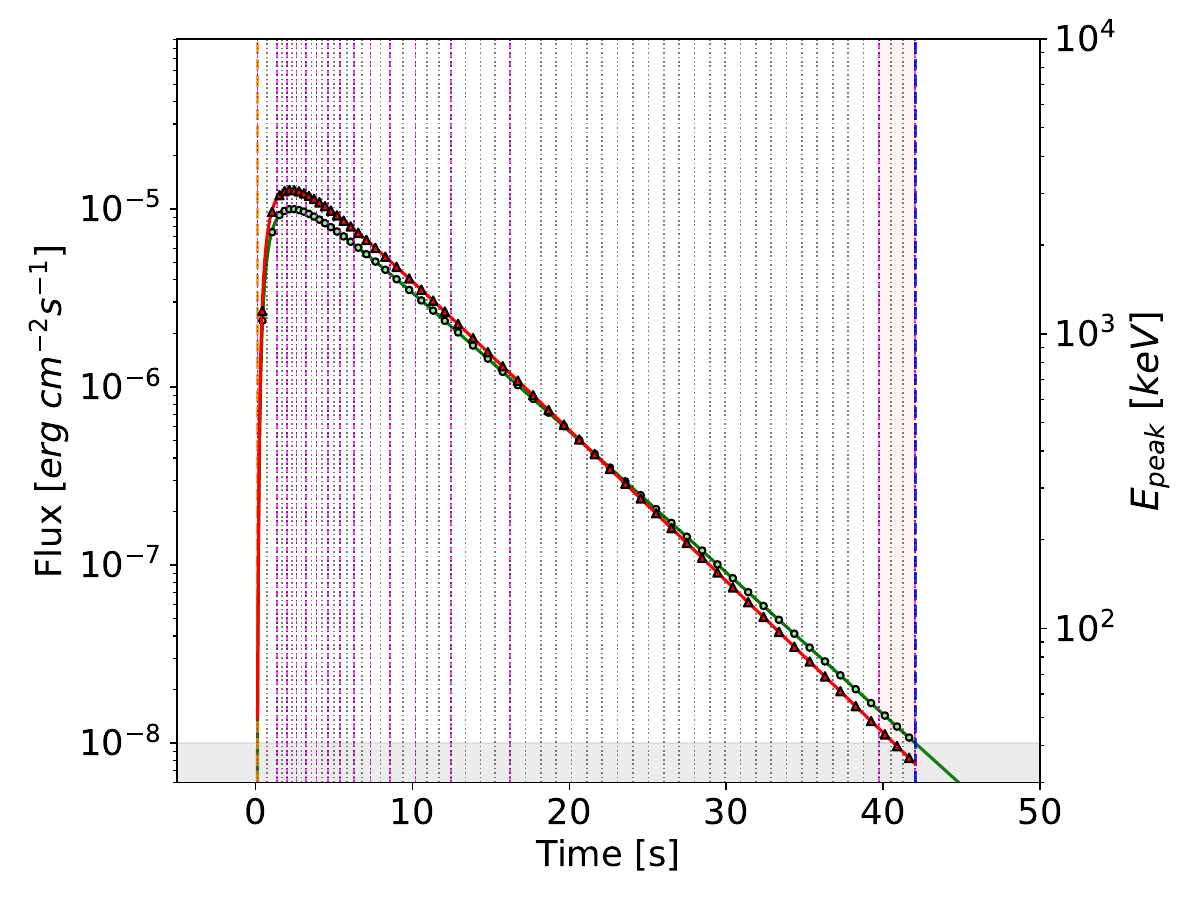}
\end{minipage}
\hspace{0.01\textwidth}
\begin{minipage}{0.47\textwidth}
\setlength{\unitlength}{1cm}
\begin{picture}(0,0)
    \put(-0.37,6){\small\textbf{(d)}}
\end{picture}
\includegraphics[width=\linewidth]{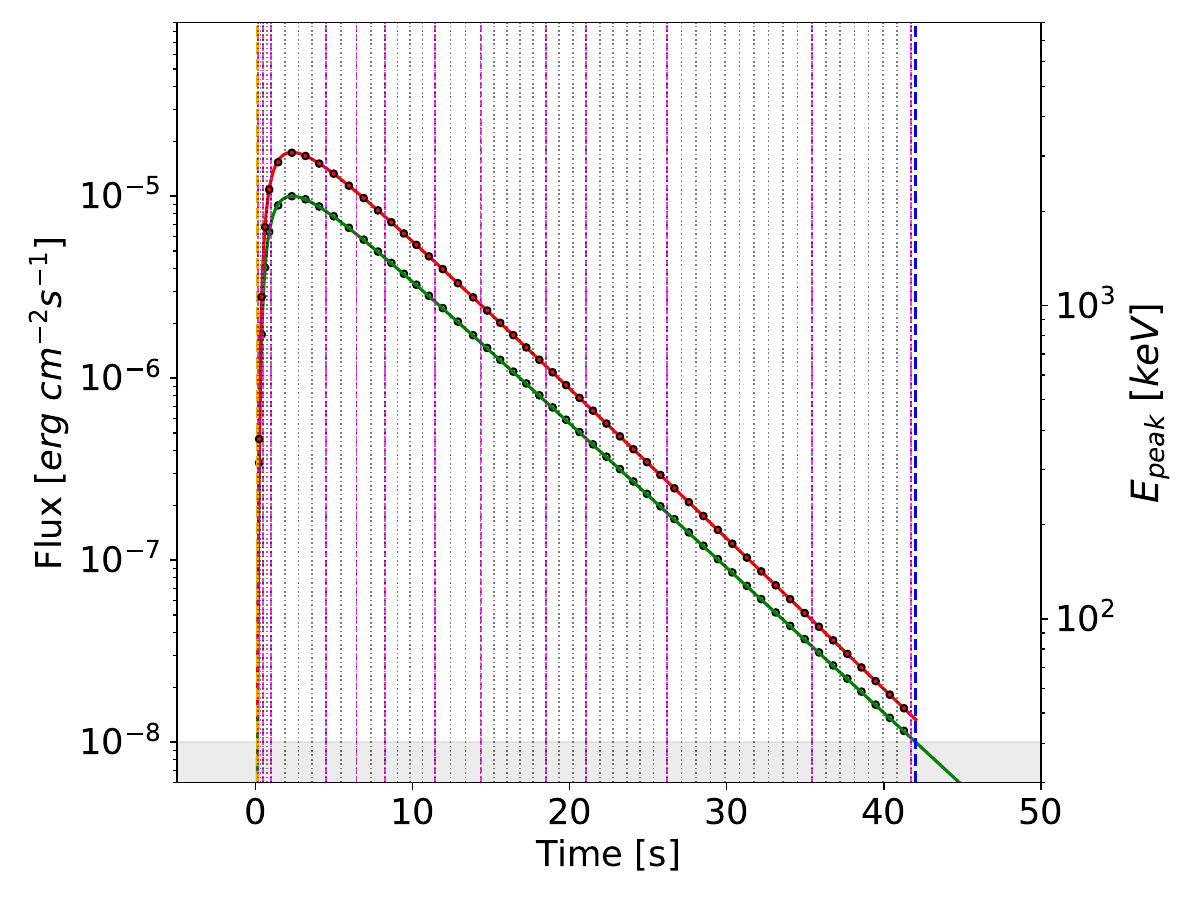}
\end{minipage}

\caption{Illustration of the temporal binning scheme and the corresponding $E_{\rm peak}$ evolution used in this study.
    Panels (a) and (b) show the hard-to-soft $E_{\rm peak}$ evolution case, while panels (c) and (d) correspond to the intensity-tracking $E_{\rm peak}$ evolution.
    In each panel, the green solid curve represents the underlying Norris pulse used to model the flux evolution and the red solid curve represents $E_{\rm peak}$ evolution.
    The vertical magenta dashed lines indicate the edges of the main time-resolved bins obtained using the two binning methods (constant fluence: (a), (c) and Bayesian blocks: (b), (d) ), whereas the thin grey vertical dotted lines mark the fine bins introduced within each main bin.
    The green points denote the average flux in each fine bin, and the red points indicate the corresponding average $E_{\rm peak}$ values.
    The red shaded regions represent time intervals where the flux falls below the adopted detector threshold and are excluded from the analysis.}
\label{fig:Flux_Epeak_Temp_Binning}
\end{figure}
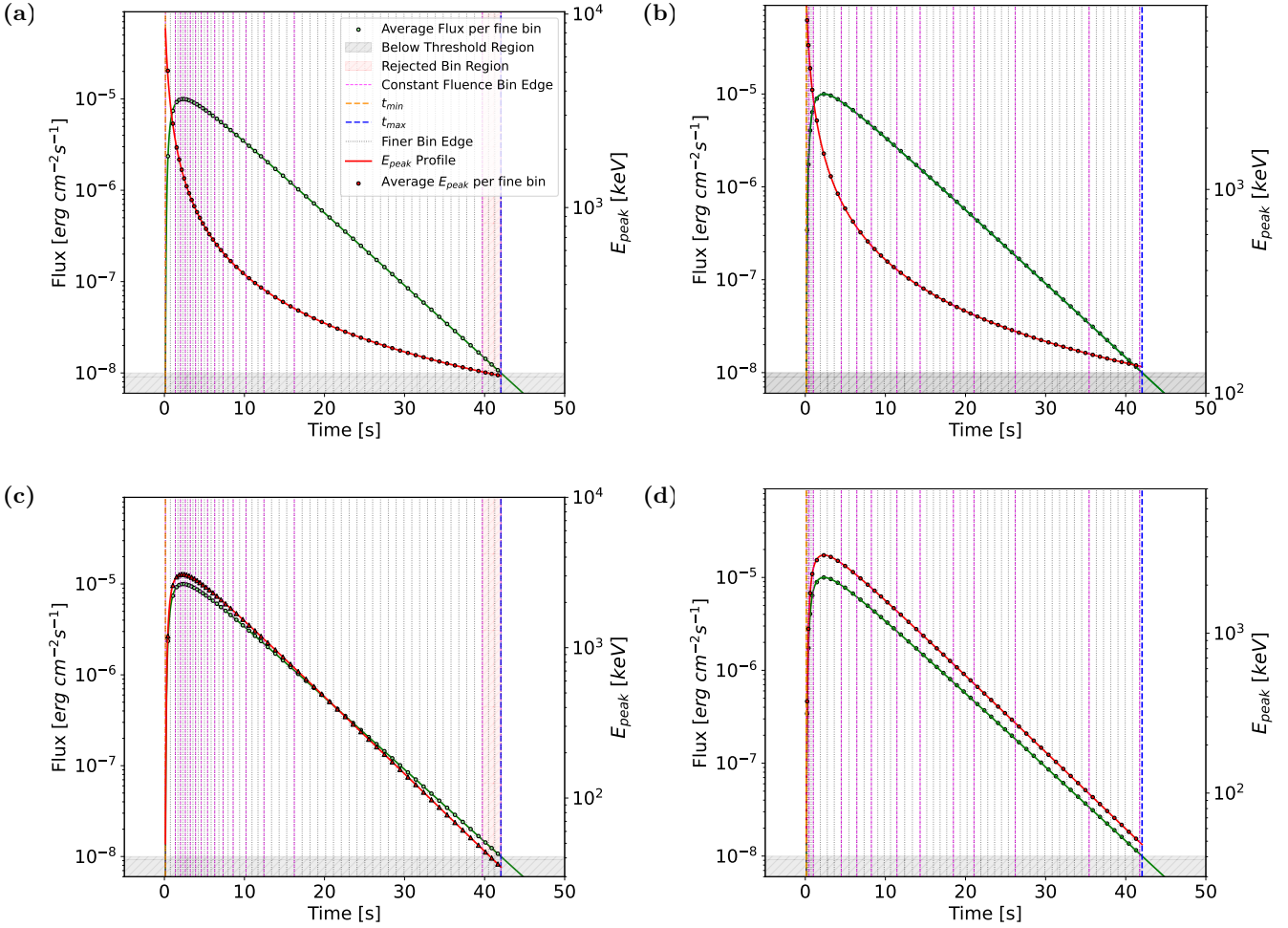

\textbf{\textit{Hard-to-soft case:}} We mathematically model the hard-to-soft evolution of the $E_{ peak}$ by a decaying power law function as given below : 
\begin{equation}
    E_{\rm peak} = a(1 + t/b)^c
    \label{Epeak_function}
\end{equation}
where $a,b,c$ are the model parameters and $t$ represents time in seconds. The maximum value of $E_{\rm peak}$ in its temporal evolution is selected based on the $E_{\rm peak}$--$\Gamma$ correlation given below, as reported by\citealt{Ghirlanda_2018}. 
\begin{equation}
\log\left(\frac{E_{\rm peak}}{300~\mathrm{keV}}\right)
=
m \log\left(\frac{\Gamma_0}{100}\right) + q \, .
\label{Gamma_Epeak_correlation}
\end{equation}
where $m = 1.28 \pm 0.03$ and $q = -0.40 \pm 0.03$. 
During the hard-to-soft phase of the temporal evolution, the maximum bulk Lorentz factor is assumed to be $\Gamma = 1000$, motivated by previous studies \citep{Iyyani2013,Iyyani2016}. Using the correlation 
described above, this corresponds to a maximum $E_{\rm peak}$ of $5290~\mathrm{keV}$. The parameters of Equation (\ref{Epeak_function}), namely $a$, $b$, and $c$, 
are therefore chosen such that the function reaches this maximum value in the beginning of the burst duration of the Norris pulse,
while ensuring that the minimum $E_{\rm peak}$ remains above $50~\mathrm{keV}$ as shown in Figure \ref{fig:Flux_Epeak_Temp_Binning}(a) and (b). The lower bound on $E_{\rm peak}$ is motivated by the results of \citealt{Preece_etal_2016}, 
who demonstrated that reliable recovery of the low-energy spectral index $\alpha$, free from observational window edge effects, is achieved when $E_{\rm peak} \gtrsim 
20~\mathrm{keV}$. We therefore adopt a conservative minimum value of $E_{\rm peak} = 50~\mathrm{keV}$ throughout this analysis. \\

\textbf{\textit{Intensity tracking case:}} In this case, the temporal evolution of $E_{\rm peak}$ is assumed to follow an intensity- (or flux-) tracking behaviour. Accordingly, we model the $E_{\rm peak}$ evolution using the Norris function, adopting parameter values similar 
to those used for the flux evolution. Only minimal adjustments are applied to ensure that the minimum $E_{peak}$ remains above $50~\mathrm{keV}$ and that the peak of the $E_{\rm peak}$ evolution 
coincides with the peak of the flux evolution, as illustrated in Figure~\ref{fig:Flux_Epeak_Temp_Binning}(c) and (d). The peak value of $E_{\rm peak}$ is determined using the correlation given in Equation~(\ref{Gamma_Epeak_correlation}), evaluated at $\Gamma = 
744$. This value of $\Gamma$ is obtained from the temporal evolution of the bulk Lorentz factor at the epoch corresponding to the maximum of the $E_{\rm peak}$ evolution.
\\
\\

{\bf (III) Temporal evolution of $\Gamma$ :}
\label{Gamma_sec}
The bulk Lorentz factor, $\Gamma$, inferred from observational signatures of non-dissipative photospheric emission in GRBs, is typically found to decrease 
monotonically with time \citep{Iyyani2016}. Observationally inferred values of $\Gamma$ generally span the range $\sim 100$--$1000$. To remain roughly 
consistent with these constraints, we adopt an upper limit of $\Gamma = 1000$ and a conservative lower limit of $\Gamma = 70$. 

\begin{figure}[htbp]
\centering

\begin{minipage}{0.45\textwidth}
\setlength{\unitlength}{1cm}
\begin{picture}(0,0)
    \put(-0.36,5.5){\small\textbf{(a)}}
\end{picture}
\includegraphics[width=\linewidth]{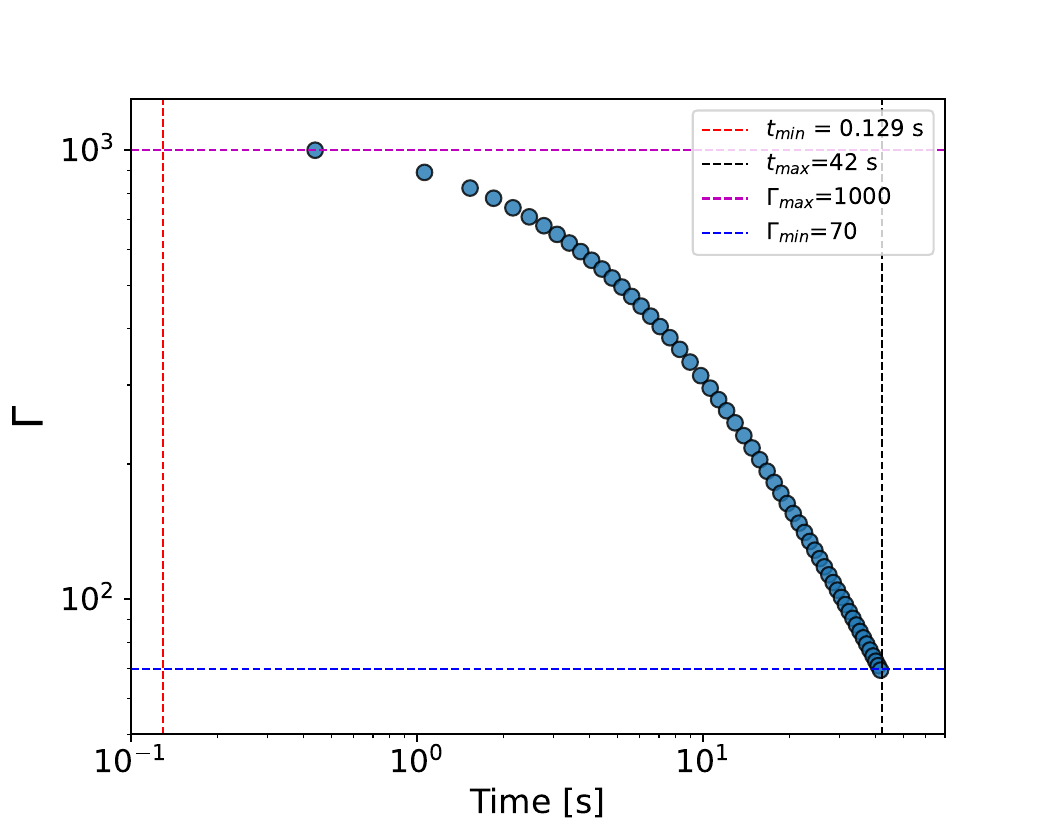}
\end{minipage}
\hspace{0.01\textwidth}
\begin{minipage}{0.45\textwidth}
\setlength{\unitlength}{1cm}
\begin{picture}(0,0)
    \put(-0.38,5.5){\small\textbf{(b)}}
\end{picture}
\includegraphics[width=\linewidth]{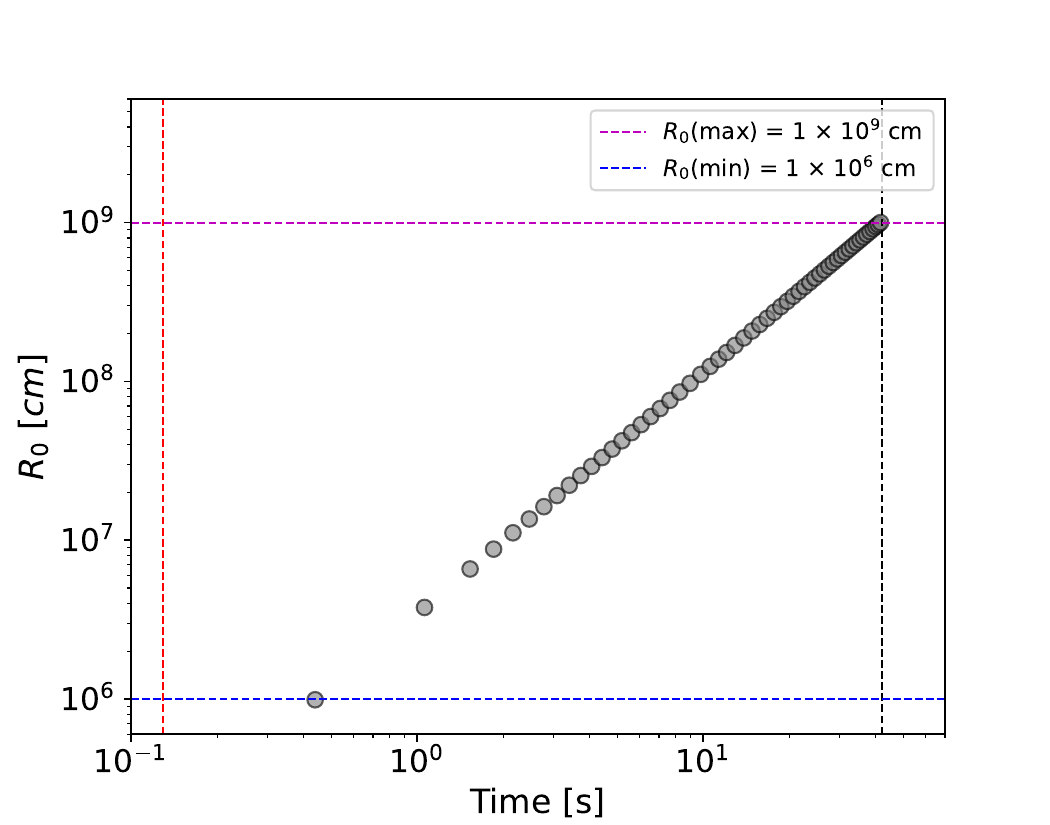}
\end{minipage}

\vspace{0.3cm}

\begin{minipage}{0.45\textwidth}
\setlength{\unitlength}{1cm}
\begin{picture}(0,0)
    \put(-0.36,5.5){\small\textbf{(c)}}
\end{picture}
\includegraphics[width=\linewidth]{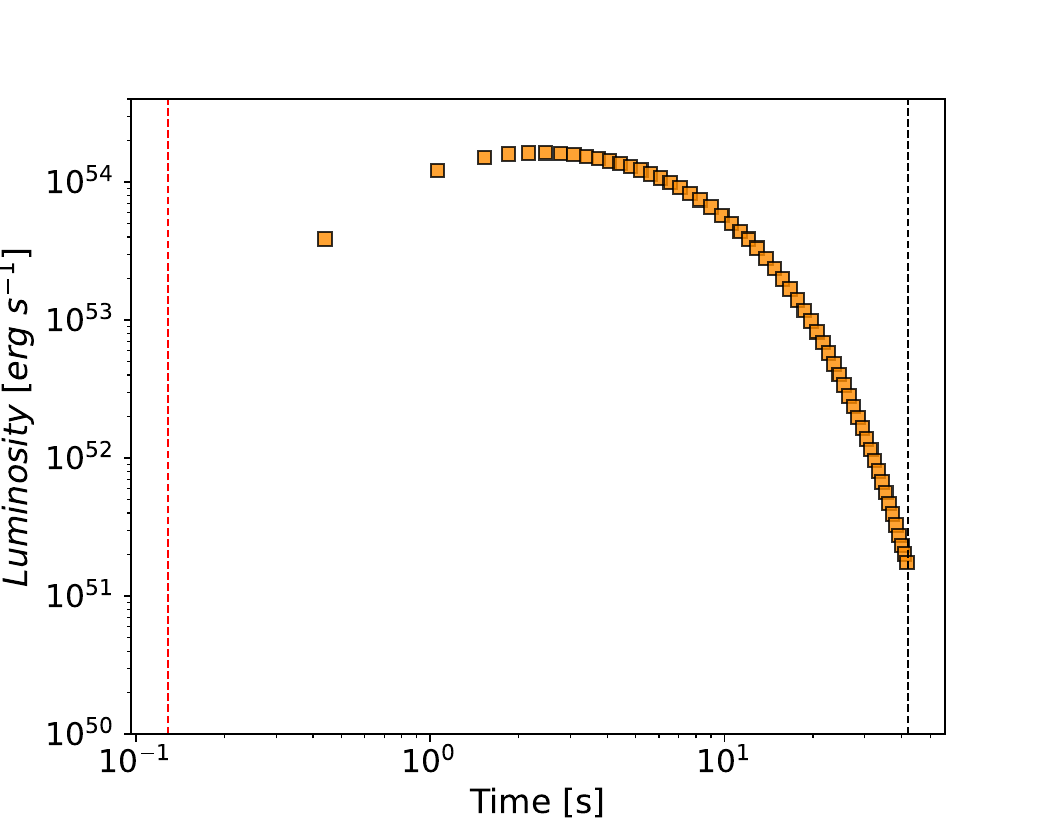}
\end{minipage}
\hspace{0.01\textwidth}
\begin{minipage}{0.45\textwidth}
\setlength{\unitlength}{1cm}
\begin{picture}(0,0)
    \put(-0.38,5.5){\small\textbf{(d)}}
\end{picture}
\includegraphics[width=\linewidth]{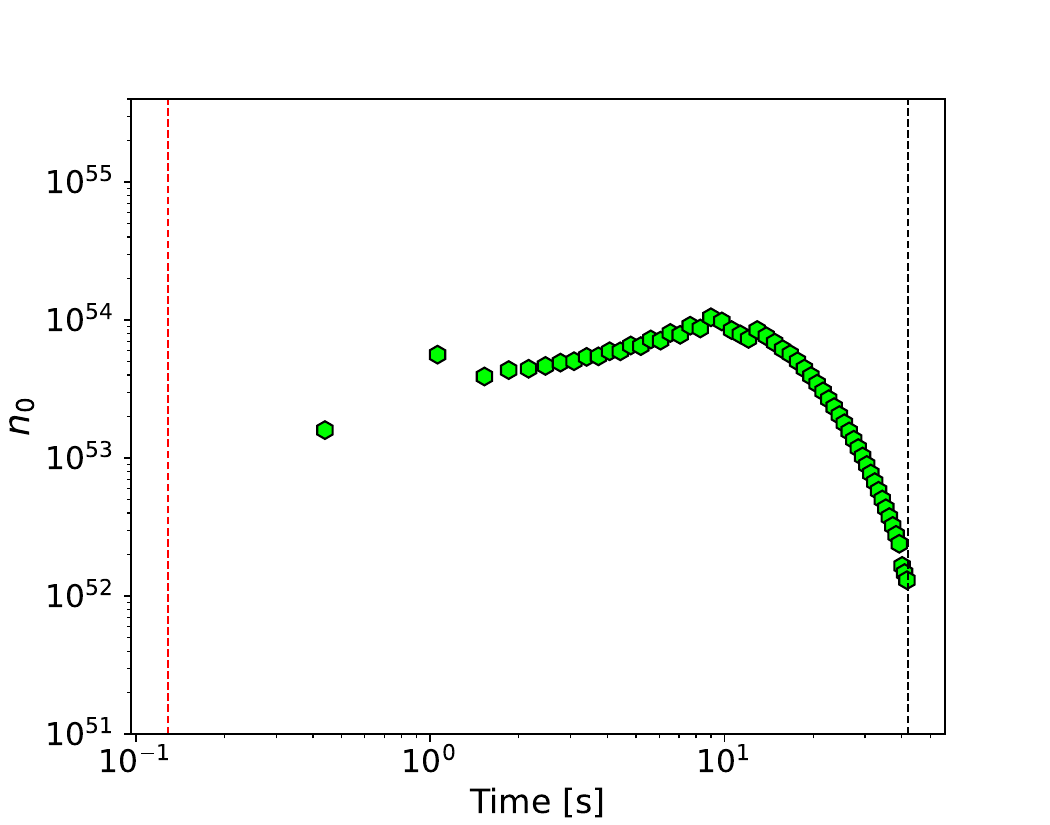}
\end{minipage}

\vspace{0.3cm}
\begin{minipage}{0.45\textwidth}
\setlength{\unitlength}{1cm}
\begin{picture}(0,0)
    \put(-0.36,5.5){\small\textbf{(e)}}
\end{picture}
\includegraphics[width=\linewidth]{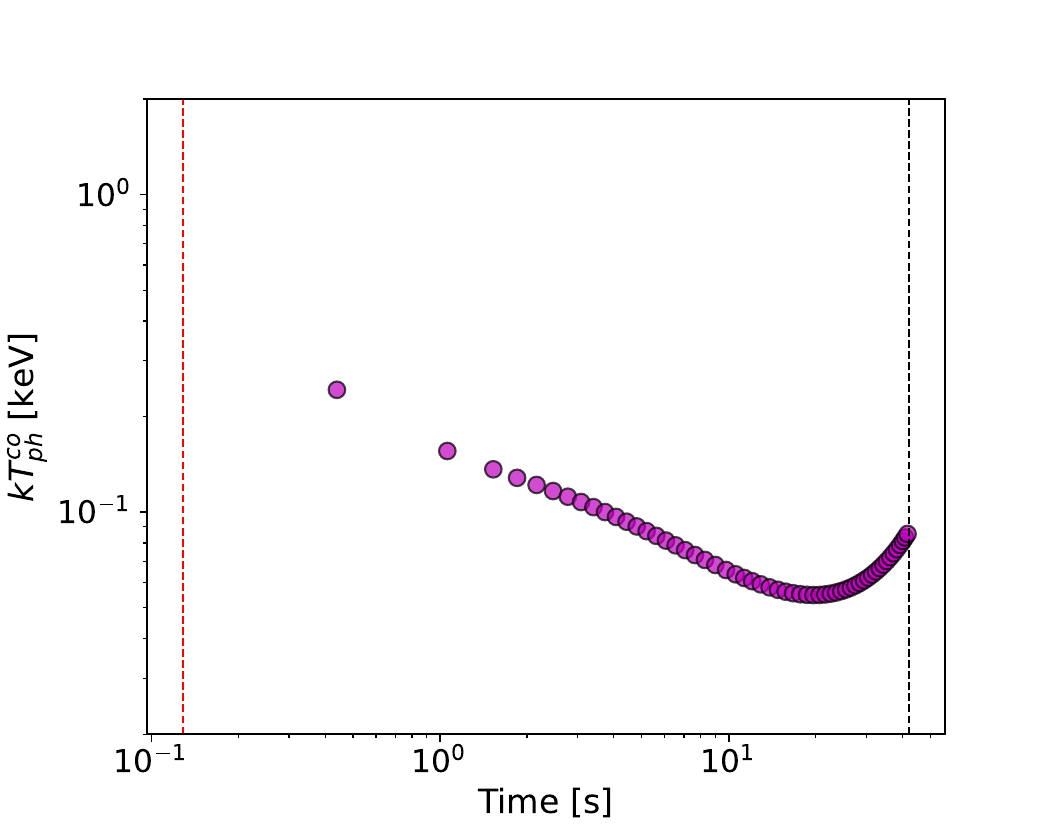}
\end{minipage}
\hspace{0.01\textwidth}
\begin{minipage}{0.45\textwidth}
\setlength{\unitlength}{1cm}
\begin{picture}(0,0)
    \put(-0.38,5.5){\small\textbf{(f)}}
\end{picture}
\includegraphics[width=\linewidth]{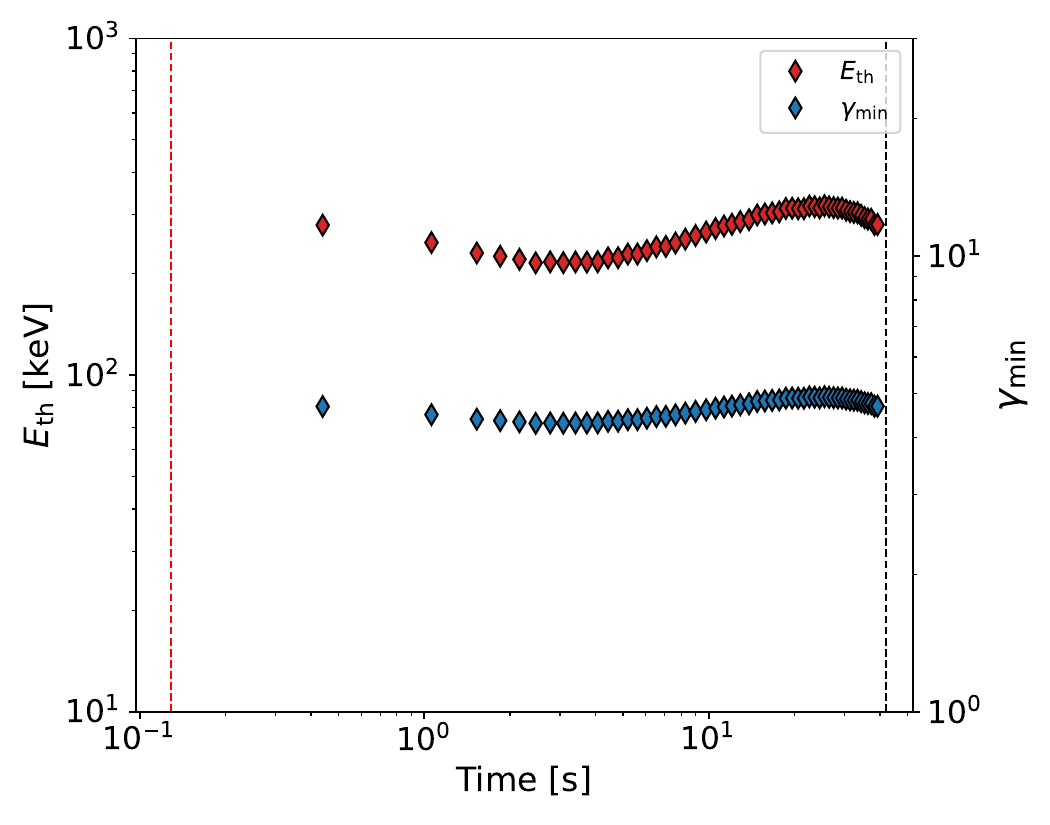}
\end{minipage}
\caption{The temporal evolution of the underlying physical parameters the for hard-to-soft $E_{\rm peak}$ evolution case are shown. (a) Bulk Lorentz factor, $\Gamma$. The magenta and blue dashed horizontal lines represent the maximum and minimum values of $\Gamma$; (b) Nozzle radius, $R_0$. The magenta and blue dashed horizontal lines represent the maximum and minimum values of $R_0$; (c) Luminosity, $L$; (d) Electron distribution normalisation, $n_0$, (e) Co-moving temperature at the photosphere, $kT_{\rm ph}^{\rm co}$ (in energy unit) and (f) electron thermal energy, $E_{th}$ (red diamonds) and the minimum Lorentz factor of power law electrons, $\gamma_{min}$ (blue diamonds), are shown.}
\label{phys_params_1}
\end{figure}

\begin{figure}[htbp]
\centering
\begin{minipage}{0.47\textwidth}
\setlength{\unitlength}{1cm}
\begin{picture}(0,0)
    \put(-0.36,5.7){\small\textbf{(a)}}
\end{picture}
\includegraphics[width=\linewidth]{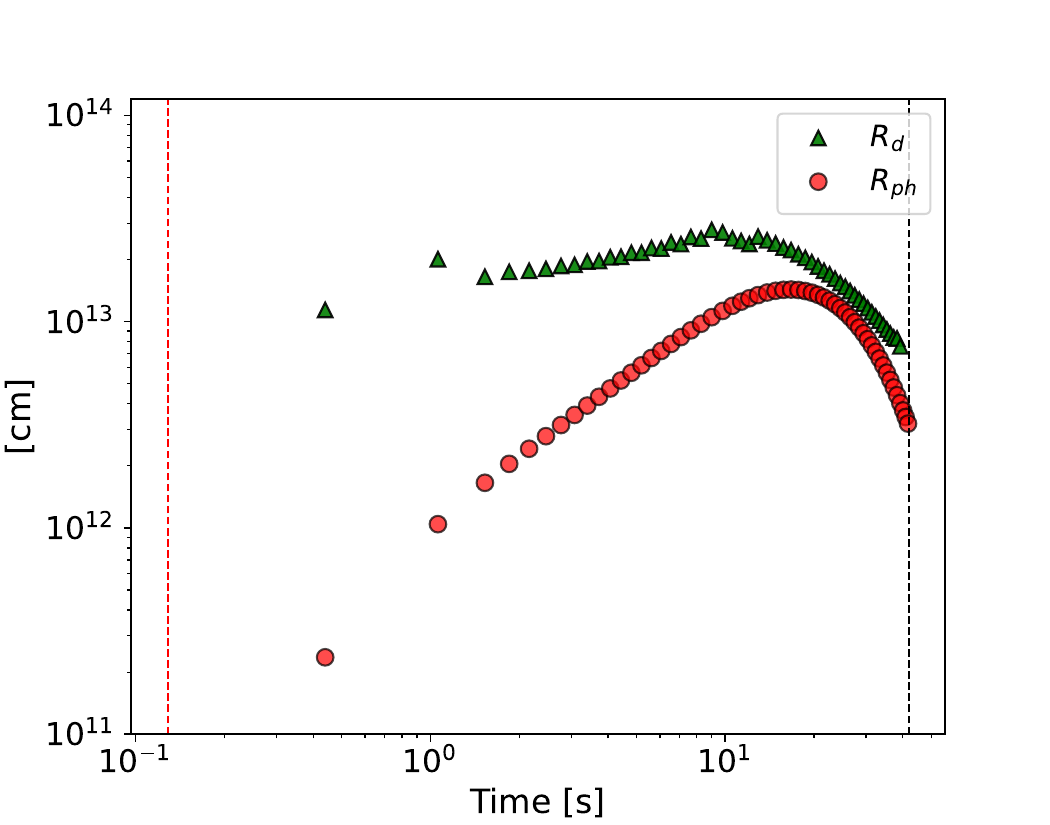}
\end{minipage}
\hspace{0.01\textwidth}
\begin{minipage}{0.47\textwidth}
\setlength{\unitlength}{1cm}
\begin{picture}(0,0)
    \put(-0.38,5.7){\small\textbf{(b)}}
\end{picture}
\includegraphics[width=\linewidth]{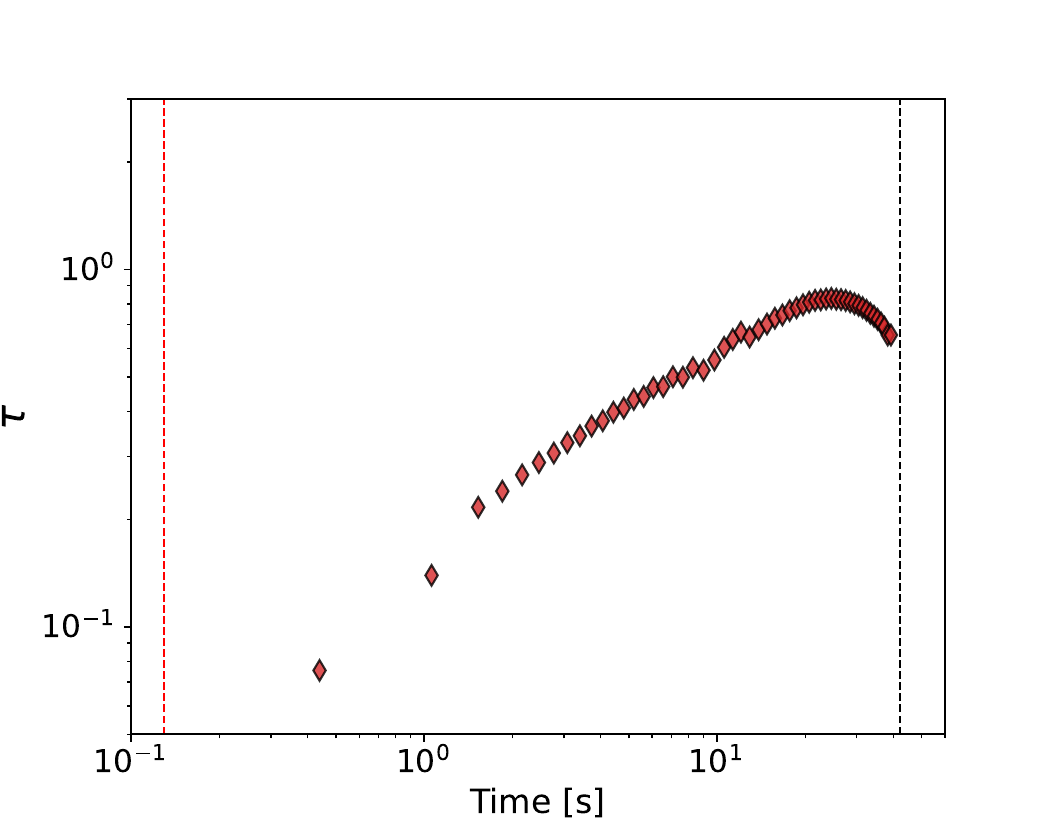}
\end{minipage}
\caption{The temporal evolution of the underlying physical parameters (a) dissipation radius, $R_d$ (green triangles) and photospheric radius $R_{ph}$ (red circles), and (b) optical depth, $\tau$ at the dissipation site are shown.}
\label{phys_params_2}
\end{figure}
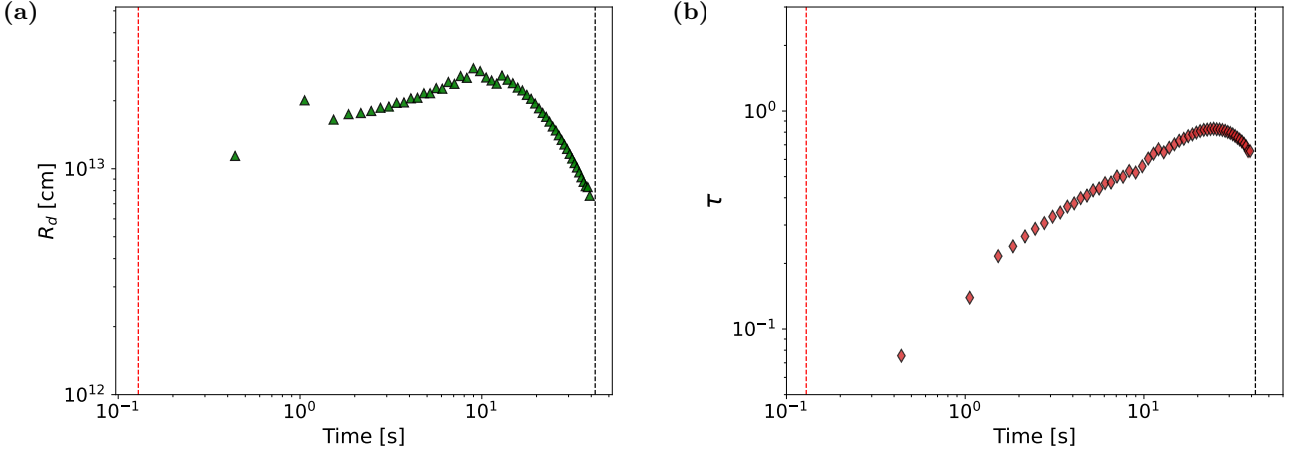

To model the temporal evolution of $\Gamma$, we adopt the decaying power-law function, similar to that given in Equation (\ref{Epeak_function}). The parameters of this function are chosen as $a = 1090$, $b = 7$, and $c = -1.42$, such that the resulting evolution naturally spans the 
desired range.
The temporal evolution plot for $\Gamma$ is shown in the Figure \ref{phys_params_1}(a).
\\
\\
{\bf (IV) Temporal evolution of nozzle radius $R_0$ :}
\label{Nozzle_sec}
The observational studies indicate that the nozzle radius ($R_0$) generally exhibits a monotonically increasing 
temporal evolution, with typical values spanning the range $10^{6}$--$10^{9}$~cm, as reported by \citealt{Iyyani2016}. Motivated by these findings, we model the temporal evolution of $R_0$ using simple power 
law function ($R_0 = norm \times (t/t_{pivot})^{k}$) but with a positive index.  The parameters of this function are chosen as $norm = 1.3 \times 10^9$ cm, $k = 1.5$ and $ t_{pivot}\ = 100$ s such that $R_0$ varies between a minimum value of 
$10^{6}$~cm and a maximum value of $10^{9}$~cm. The resulting temporal evolution of $R_0$ is shown in Figure~\ref{phys_params_1}(b).
\\
\\
{\bf (V) Temporal evolution of luminosity $L$ :}
We assume a redshift of $z = 1$, corresponding to a luminosity distance of $d_L = 6.8 \times 10^6 \, \ \rm kpc$, calculated under the assumption of a flat $\Lambda$CDM cosmology with cosmological parameters adopted from the \textit{Planck} Collaboration results \citep{Planck_2020}, including a Hubble constant of $H_0 = 67.4 \, \ \rm km\,s^{-1}\,Mpc^{-1}$. The isotropic-equivalent luminosity $L$ is related to the observed flux $F_{\rm ob}$ and the luminosity distance through
\begin{equation}
L = 4\pi d_L^2\, Y\, F_{\rm ob},
\end{equation}
where $Y$ is the inverse of the radiative efficiency and is taken to be $Y = 5$ \citep{Racusin2011}, an average value observed for long duration GRBs detected by {\it Fermi} GBM. Since the observed flux, $F_{\rm ob}$, is modeled using the Norris pulse profile, the adopted values of $z$, $d_L$, and $Y$ uniquely determine the temporal evolution of the burst luminosity, $L$. The resulting luminosity evolution is presented in Figure~\ref{phys_params_1}(c).
\\
\\
{\bf (VI) Temporal evolution of electron normalization $n_0$ :}
\label{Elec_norm}
The total number of electrons contained within the volume element traversed by the outflow over an observational timescale $t_{\rm obs}$ can be estimated as
\begin{equation}
N_e \sim \frac{L}{\Gamma\, m_p c^2}\, t_{\rm obs},
\label{Electron_num}
\end{equation}
where $m_p$ denotes the proton mass and $c$ is the speed of light. The normalization parameter of the electron distribution, $n_0$, is taken to be proportional to $N_e$. 
Consequently, by using the temporally evolving values of the bulk Lorentz factor $\Gamma$ and the luminosity $L$, together with the observational timescale $t_{\rm obs}$, 
we derive the temporal evolution of $n_0$, as shown in Figure~\ref{phys_params_1}(d). Here, $t_{\rm obs}$ corresponds to the fine-bin widths adopted in Section~\ref{binning}.
\\
\\
{\bf (VII) Temporal evolution of photospheric radius $R_{ph}$ :}
\label{Photo_section}
The photospheric radius $R_{ph}$ depends on the luminosity $L$ and the bulk Lorentz factor $\Gamma$, as 
given by the equation (\ref{Photo_radius_1}). Since the temporal evolution of both $L$ and $\Gamma$ has already been specified, the corresponding temporal evolution of $R_{ph}$ can be directly determined. \textbf{The resulting evolution of $R_{ph}$ is shown in the Figure \ref{phys_params_2} and Figure \ref{phys_params_4}. }
\\
\\
{\bf (VIII) Temporal evolution of comoving photospheric temperature $kT_{\rm ph}^{\rm co}$ :}
\label{Seed_temp}
As discussed in Section~\ref{Input_para}, the comoving photospheric temperature $kT_{\rm ph}^{\rm co}$ \ depends on the luminosity $L$, the bulk Lorentz factor $\Gamma$, and the nozzle 
radius $R_0$. Since the temporal evolution of these quantities has already been specified, we compute the corresponding time-dependent values of the temperature of 
the thermal emission in energy scale at the photosphere in the co-moving frame, $kT_{\rm ph}^{\rm co}$. The resulting evolution of $kT_{\rm ph}^{\rm co}$ is shown in Figure~\ref{phys_params_1}(e).
\\
\\
{\bf (IX) Temporal evolution of electron's thermal temperature $E_{th}$ :}
In the optically thin inverse Compton scattering regime, the shape of the Comptonised spectrum closely reflects the 
underlying electron energy distribution \citep{Bordoloi_Iyyani_2025}. Consequently, for a hybrid electron distribution, the resulting spectrum is expected to exhibit a distinct spectral hump produced by 
the thermal electron population, along with an extended high-energy power-law tail arising from the accelerated electrons.

The location of the Comptonized spectral hump is primarily determined by the seed photon temperature, $kT_{\rm ph}^{\rm co}$, and the characteristic thermal electron energy, $E_{\rm th}$. The observed spectral peak, $E_{\rm peak}$, corresponds to the Doppler-boosted Comptonized peak in the observer frame and is prescribed through Equation~(\ref{Epeak_function}) in Item~{\bf II}. The temporal evolution of the seed photon temperature, $kT_{\rm ph}^{\rm co}$, has already been defined in Item~{\bf VIII}.  

Therefore, in the ICS spectral simulations described in Section~\ref{Input_para}, we determine the values of $E_{\rm th}$ required to reproduce Comptonized spectra with the prescribed observer-frame $E_{\rm peak}$. Using the adopted temporal evolution of $E_{\rm peak}$ together with the observed evolution of $kT^{\rm obs}$, we derive the corresponding time-dependent evolution of $E_{\rm th}$. The resulting evolution profiles are shown in Figure~\ref{phys_params_1}(f) and Figure~\ref{phys_params_3}(f) of Appendix~\ref{Input_parameter_intensity_tracking}.

A key difference between the hard-to-soft and intensity-tracking scenarios emerges after $\sim 20$~s. In the intensity-tracking case, $E_{\rm th}$ decreases steadily with time, whereas in the hard-to-soft case it remains nearly constant. Since the seed photon temperature, $kT_{\rm ph}^{\rm co}$, evolves identically in both scenarios, this difference is driven primarily by the distinct temporal evolution patterns of $E_{\rm peak}$. In particular, the hard-to-soft evolution of prescribed $E_{\rm peak}$ becomes comparatively flatter after $\sim 20$~s, leading to a correspondingly weaker evolution in $E_{\rm th}$.\\

The minimum Lorentz factor of electrons in the power-law tail of the distribution, $\gamma_{\min}$, is related to the electron thermal energy through
\begin{equation}
E_{\rm th} = \left(\frac{\gamma_{\min}}{3} - 1\right) m_e c^2,
\end{equation}
where $\gamma_{\min} = 3\,\gamma_{\rm th}$ and $\gamma_{\rm th}$ corresponds to the average electron Lorentz factor of the thermal part of the electron distribution. The temporal evolution of $\gamma_{\min}$ is shown by the blue diamonds in Figures~\ref{phys_params_1}(f) and \ref{phys_params_3}(f). For the hard-to-soft $E_{\rm peak}$ evolution case, $\gamma_{\min}$ remains nearly constant around a value of $\sim 5$. In contrast, for the intensity-tracking $E_{\rm peak}$ evolution case, it varies between $\sim 3.2$ and $6.2$. Notably, these values indicate that $\gamma_{\min}$ lies in the mildly relativistic regime. 
\\
\\
{\bf (X) Temporal evolution of the dissipation radius $R_d$ :}
The flux of the Comptonised spectrum depends on both the seed photon flux and the normalization of the electron distribution \citep{rybicki}. 
The observed energy-integrated flux evolution, modeled using the Norris pulse (equation \ref{Norris_fnc}), represents the Doppler-boosted flux of the Comptonised spectrum produced in the comoving frame. The normalization of the electron distribution, $n_0$, has been specified accordingly, and the seed photon temperature $kT_{\rm ph}^{\rm co}$ is fixed as described above.

The dissipation radius, $R_d$, is determined by requiring the Doppler-boosted inverse Compton scattering (ICS) spectrum to reproduce the prescribed prompt-emission flux evolution described by the Norris pulse. The seed photons originate from the photospheric thermal emission, with $R_{\rm ph}$ and $kT_{\rm ph}^{\rm co}$ determined self-consistently from the adopted luminosity, $\Gamma$, and $R_0$, as described in Section~\ref{Photo_section}. As the seed photons propagate from $R_{\rm ph}$ to $R_d$, their flux is geometrically diluted as $(R_{\rm ph}/R_d)^2$, while in the optically thin regime a fraction $\tau$ undergoes first-order IC scattering (Section \ref{Input_para}). Thus, the seed-photon flux available for scattering scales as
\begin{equation}
F_{\rm seed}(R_d)
=
F_{R_{\rm ph}}
\left(\frac{R_{\rm ph}}{R_d}\right)^3,
\end{equation}
where $F_{R_{\rm ph}}$ is the photospheric flux at $R_{ph}$. For the adopted electron normalization $n_0$, $R_d$ is then obtained by requiring the resulting Doppler-boosted ICS spectrum to reproduce the prescribed total prompt-emission flux. Hence, neither $R_d$ nor the optical depth $\tau$ is a free parameter; the latter follows from
\begin{equation}
\tau=\frac{R_{\rm ph}}{R_d}.
\end{equation}

The resulting temporal evolution of $R_d$ and $\tau$ is shown in Figure~\ref{phys_params_2}(a) and (b), respectively. Throughout the burst evolution, $R_d$ remains of order $\sim10^{13}$~cm and $R_d>R_{\rm ph}$, giving $\tau\lesssim1$. Thus, confirming that the simulated spectra are consistently produced within the optically thin ICS regime.
\\
\\
Overall, Figures~\ref{phys_params_1} and \ref{phys_params_2} summarize the underlying physical parameter evolution adopted to generate a GRB emission pulse exhibiting hard-to-soft spectral evolution. A corresponding set of parameter evolutions used for the intensity-tracking scenario is provided in Appendix~\ref{Input_parameter_intensity_tracking}.

\subsection{Simulation of the ICS spectrum}
\label{Simul_spec}
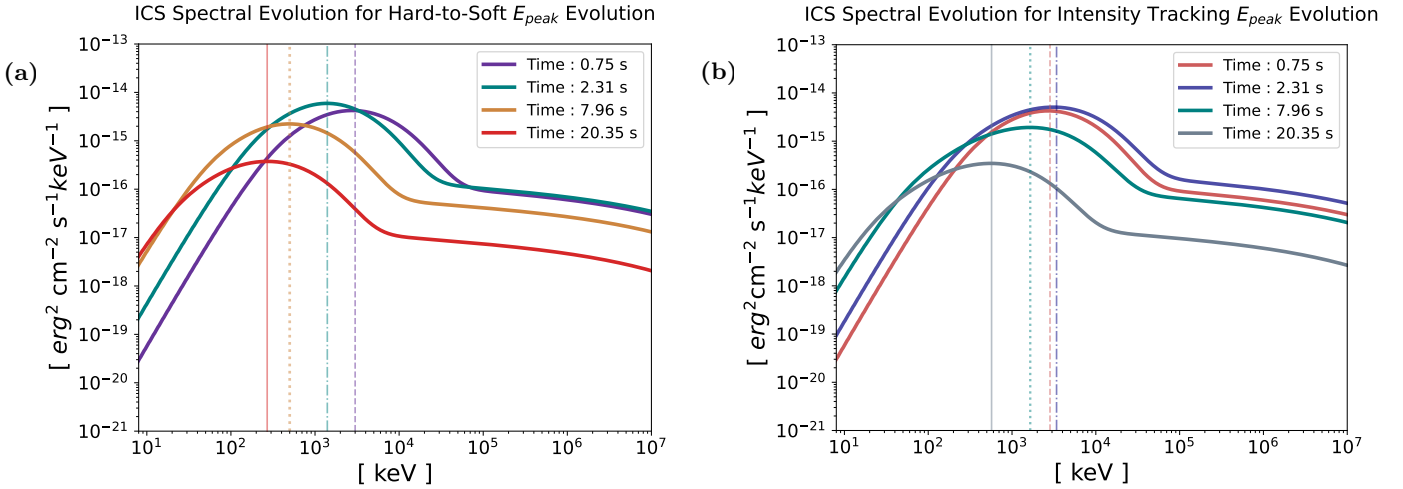
\begin{figure}[htbp]
\centering

\begin{minipage}{0.47\textwidth}
\setlength{\unitlength}{1cm}
\begin{picture}(0,0)
    \put(-0.36,5.5){\small\textbf{(a)}}
\end{picture}
\includegraphics[width=\linewidth]{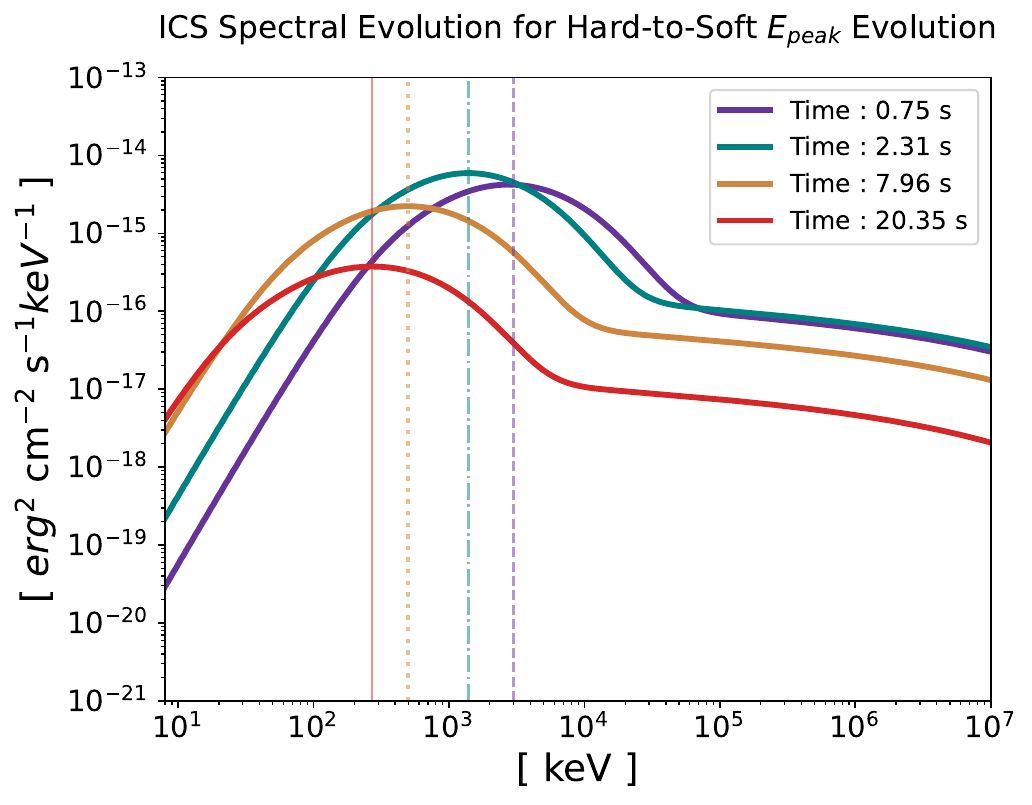}
\end{minipage}
\hspace{0.03\textwidth}
\begin{minipage}{0.48\textwidth}
\setlength{\unitlength}{1cm}
\begin{picture}(0,0)
    \put(-0.38,5.5){\small\textbf{(b)}}
\end{picture}
\includegraphics[width=\linewidth]{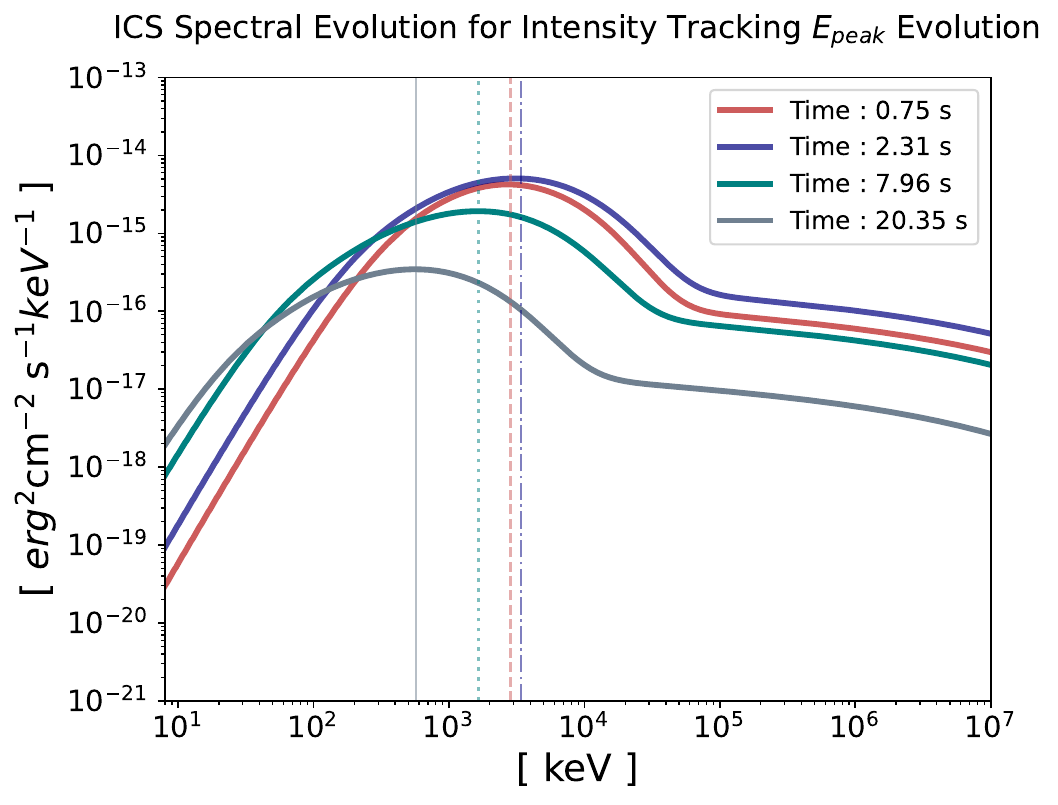}
\end{minipage}

\vspace{0.3cm}

\caption{A representative simulated ICS spectra at different epochs of the GRB emission pulse, boosted to the observer frame. The solid curves show the spectra, while the same-colored vertical lines mark the corresponding $E_{\rm peak}$ values. In panel (a), representing the hard-to-soft evolution case, the dashed, dash-dotted, dotted, and solid vertical lines correspond to $E_{\rm peak} \sim 3000$ keV, 1400 keV, 500 keV, and 270 keV, respectively. In panel (b), representing the intensity-tracking case, the dashed, dash-dotted, dotted, and solid vertical lines correspond to $E_{\rm peak} \sim 3400$ keV, 2800 keV, 1700 keV, and 600 keV, respectively.}
\label{fig:sim_spec_h_t_s}

\end{figure}

Using the emission scenario and the physically motivated parameter space described in Section~\ref{Input_para} and Section~\ref{Physical_para_setup} respectively, we simulate the time-dependent inverse Compton scattering (ICS) spectra at different epochs along the Norris flux pulse profile using the {\it Naima} package.
A representative set of simulated spectra for the hard-to-soft and intensity-tracking cases are shown in Figures~\ref{fig:sim_spec_h_t_s} (a) and \ref{fig:sim_spec_h_t_s} (b) respectively. Here, we simulate the ICS spectra in the comoving frame and then boost them to the observer frame while using the corresponding predefined seed-photon temperatures, ensuring that the spectral peaks in the observer frame remain consistent with the predefined $E_{\rm peak}$ values at different epochs of the Norris flux profile.

\section{Time Resolved Spectral Analysis}
\label{Time_resol_spec_analy}
\subsection{Binning Methods Applied}
\label{binning}
To perform time-resolved spectral analysis of GRB emission pulses exhibiting hard-to-soft and intensity-tracking behaviour, the flux profiles must first be temporally resolved. We adopt two binning schemes commonly used in GRB studies: (i) constant-fluence binning, analogous to signal-to-noise ratio based binning, and (ii) Bayesian blocks binning.\\

{\bf Constant fluence binning:} In the constant-fluence binning approach, each time bin is constructed to contain a fixed fluence of $6.06 \times 10^{-6}~\mathrm{erg\,cm^{-2}}$. This results in a total of 14 time-resolved bins (hereafter referred to as the main bins), collectively covering nearly the entire 
burst duration, as indicated by the magenta dashed vertical lines in Figure~\ref{fig:Flux_Epeak_Temp_Binning}(a) and (c). The 
chosen fluence threshold minimizes the remaining time interval for which the required fluence criterion is not satisfied, thereby reducing the unbinned portion of the burst duration (shown by the red shaded region in 
Figure~\ref{fig:Flux_Epeak_Temp_Binning}). This choice maximizes the temporal coverage of the burst while ensuring a sufficient number of time-resolved bins (below $20$) for statistically meaningful correlation analysis study to be conducted. 
\\
\\
{\bf Bayesian block binning:} On the other hand, when the Bayesian blocks binning method \citep{scargle_1998} is applied to the Norris function, a total of 12 time-resolved bins are obtained, as indicated by the magenta dashed vertical lines in Figure~\ref{fig:Flux_Epeak_Temp_Binning}(b) and Figure~\ref{fig:Flux_Epeak_Temp_Binning}(d). This method defines the bin boundaries by identifying statistically significant changes in the flux as a function of time with false-positive probability $p_0 = 0.01$.\\

For each of the two $E_{peak}$ evolution patterns hard-to-soft and intensity tracking, we apply both binning techniques independently. This results in a total of four distinct cases in our study (Figure \ref{fig:Flux_Epeak_Temp_Binning}): Case 1 - Hard-to-soft $E_{peak}$ and constant fluence binning; Case 2 - Hard-to-Soft $E_{peak}$ and Bayesian Block binning; Case 3 - Intensity tracking and constant fluence binning and Case 4 - Intensity tracking and Bayesian Block binning.  Each case is analyzed separately to investigate the $E_{peak}$--$\alpha$ correlation.  \\
The time-resolved intervals (main bins) obtained from the two binning methods span durations from $\sim$0.3~s to $\sim$24~s, significantly longer than the typical cooling timescale of the inverse Compton scattering process ($\lesssim 10^{-6}$~s; \citealt{Bordoloi_Iyyani_2025}). To better capture the underlying physical and temporal evolution, each main 
bin is further subdivided into finer time bins of 1~s duration, indicated by the grey dashed vertical lines in Figure~\ref{fig:Flux_Epeak_Temp_Binning}. For main bins shorter than 1~s, the interval is divided into at least two finer bins with sub-second durations. An IC spectrum is simulated for each fine time bin, ensuring that every main bin contains at least two simulated spectra.\\

{\bf \textit{Motivation for adopting two binning methods:}}\\
The choice of time binning can influence the recovered spectral evolution and, consequently, the inferred $E_{\rm peak}$--$\alpha$ correlation. Both Bayesian-block and constant-fluence 
binning are commonly used in time-resolved GRB analyses, but they sample the underlying, pre-defined $E_{\rm peak}$ evolution differently, as illustrated in Figures~2(a) and~2(b). The Bayesian-block binning adapts the bin widths to the temporal structure of the 
pulse and therefore provides a more balanced sampling of the rising and decaying phases of the Norris pulse. In contrast, constant-fluence binning allocates narrower bins to the brighter portions of the pulse and broader bins to the fainter regions, thereby preferentially sampling the high-flux portion of the pulse. Consequently, the two schemes can result in different numbers and 
distributions of $E_{\rm peak}$-$\alpha$ data points across the rising and decaying phases. Since the inferred correlation is based on these time-resolved data points, it is therefore important to investigate whether the resulting $E_{\rm peak}$-$\alpha$ evolution and correlation are sensitive to the adopted binning scheme.
Furthermore, the different durations of the main bins lead to different numbers of fine-bin spectra being integrated in our simulations, potentially introducing different degrees of spectral smearing. This can affect the recovered $\alpha$ and $E_{\rm peak}$ 
and hence the $E_{\rm peak}$--$\alpha$ correlation. Temporal binning has previously been shown to influence the recovery of GRB spectral parameters, with Bayesian-block binning providing a more accurate 
representation of the intrinsic spectral evolution than some alternative schemes \citep{Burgess2014_1}. We, therefore, employ both binning schemes to test the robustness of the predicted $E_{\rm peak}$--$\alpha$ correlation against the choice of time-binning strategy.

\subsection{Spectral Modeling with Band function}
\label{spec_model_band}
Fake count spectra are generated within the \texttt{3ML} framework \citep{Vianello_etal_2015} by forward-folding the simulated inverse Compton (IC) photon spectra through the detector response matrices and subsequently applying Poisson fluctuations to the resulting counts. 
This procedure produces realistic synthetic observations for each fine time bin. For all four case studies, we use the same representative NaI and BGO detector response files adopted from an observed GRB 131014A. A synthetic background spectrum is generated using a power-law model with an index of $-0.5$.

\begin{figure}[htbp]
\centering


{\large\textbf{Hard-to-Soft $E_{\rm peak}$ evolution Case}}

\vspace{0.25cm}


\begin{minipage}{0.47\textwidth}
\centering
\setlength{\unitlength}{1cm}
\begin{picture}(0,0)
    \put(-4.7,-0.3){\small\textbf{(a)}}
\end{picture}
\includegraphics[width=\linewidth]{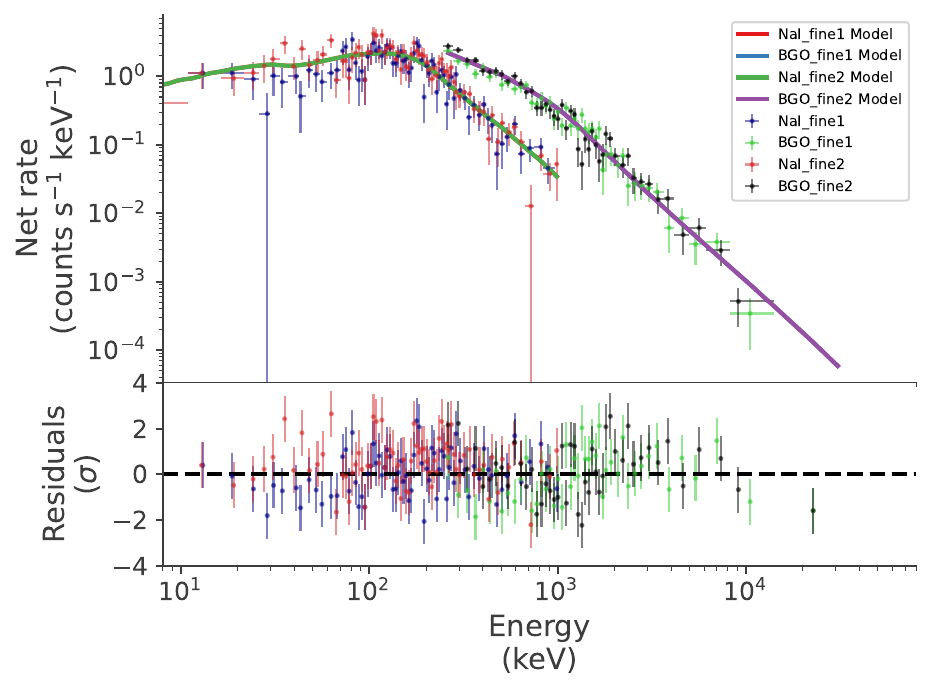}
\end{minipage}
\hspace{0.01\textwidth}
\begin{minipage}{0.42\textwidth}
\centering
\setlength{\unitlength}{1cm}
\begin{picture}(0,0)
    \put(-4.3,0.1){\small\textbf{(b)}}
\end{picture}
\includegraphics[width=\linewidth]{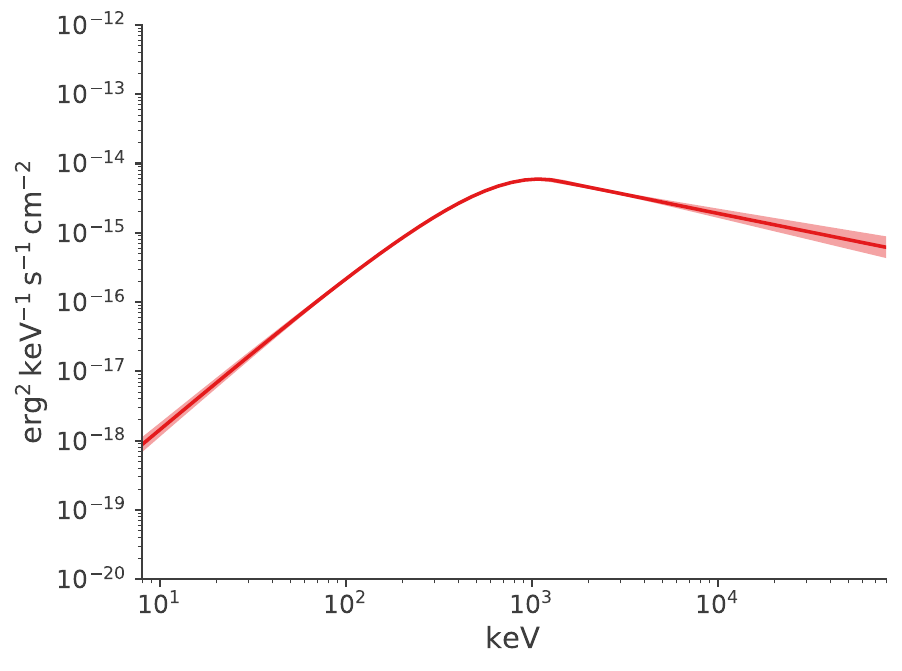}
\end{minipage}


\vspace{0.6cm}


{\large\textbf{Intensity-Tracking $E_{\rm peak}$ evolution Case}}

\vspace{0.25cm}


\begin{minipage}{0.47\textwidth}
\centering
\setlength{\unitlength}{1cm}
\begin{picture}(0,0)
    \put(-4.7,-0.3){\small\textbf{(c)}}
\end{picture}
\includegraphics[width=\linewidth]{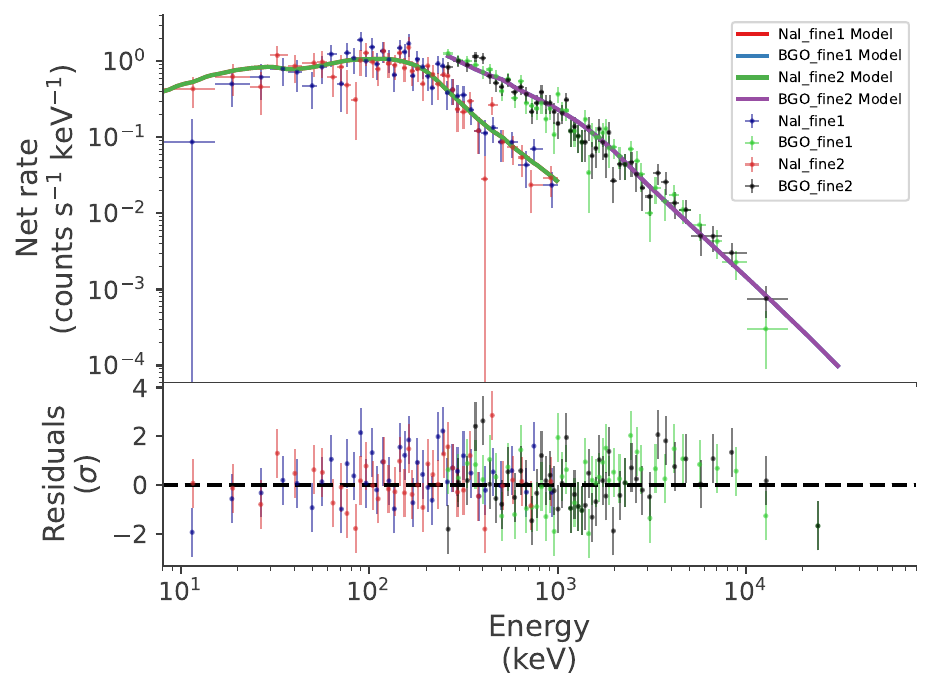}
\end{minipage}
\hspace{0.03\textwidth}
\begin{minipage}{0.42\textwidth}
\centering
\setlength{\unitlength}{1cm}
\begin{picture}(0,0)
    \put(-4.3,0.007){\small\textbf{(d)}}
\end{picture}
\includegraphics[width=\linewidth]{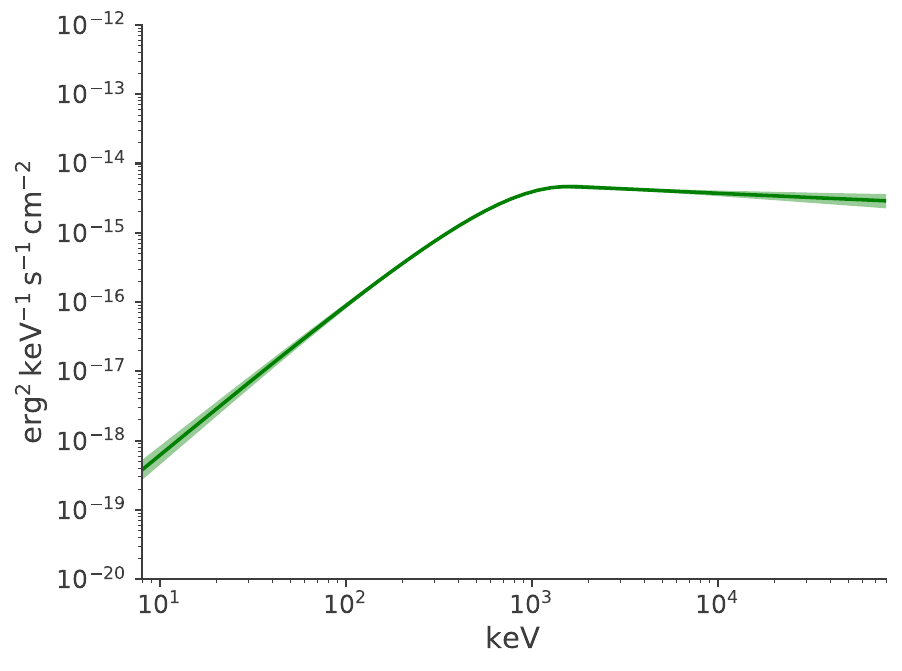}
\end{minipage}


\caption{
Counts-plus-residual plots (left panels) and corresponding $\nu F_{\nu}$ spectra (right panels) for the peak bin in the hard-to-soft and intensity-tracking spectral-evolution cases using constant-fluence binning.
}

\label{count_residual_plot_1}

\end{figure}
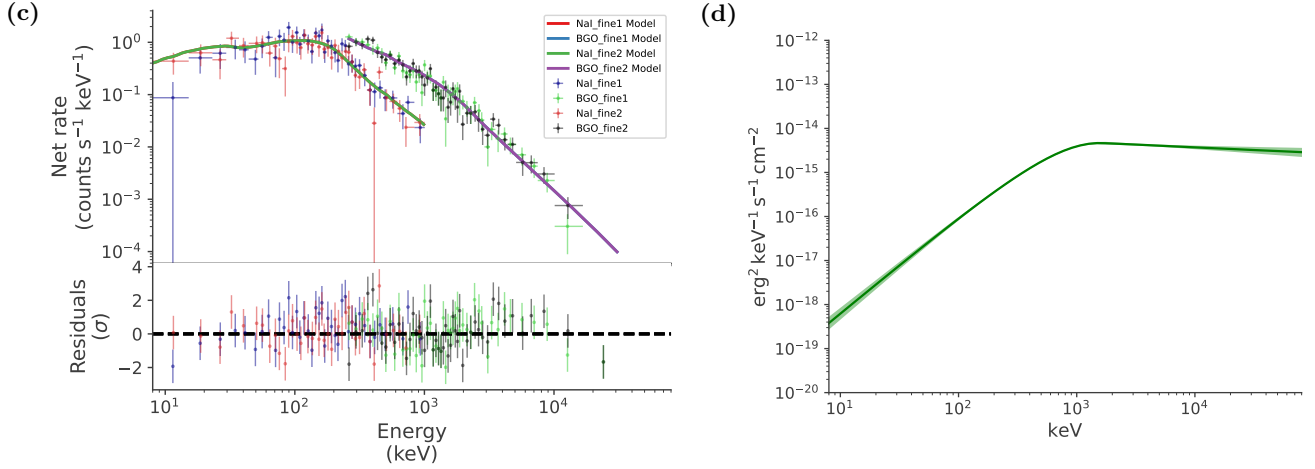

For each main time interval defined by the constant-fluence and Bayesian blocks binning schemes, the simulated count spectra corresponding to all fine bins within that interval are jointly analyzed using the Band 
function. Since Poisson fluctuations are applied to the forward-folded count spectra, each realization contains different statistical count variations, leading to corresponding fluctuations in the fitted spectral parameters. To quantify this uncertainty, the fitting procedure is repeated 250 times for each main bin, with 
the background counts independently resampled from the corresponding Poisson distribution in every realization. This produces distributions of the fitted $\alpha$ and $E_{\rm peak}$ values, from which the median values are computed and subsequently used for the correlation analysis presented in the following section.\\
The quality of the Band function fit is demonstrated using both count-plus-residual and the corresponding $\nu F_{\nu}$ plots in the peak region of the Norris flux profile for Case 1 and Case 3 in Figure \ref{count_residual_plot_1}, and for case 2 and case 4 in the Figure \ref{count_residual_plot}. A visual inspection of the residuals shows no systematic structure and a random distribution around zero, indicating that the Band function provides an adequate description of the data.

\section{Results}
\label{results}
\begin{figure}[htbp]
\centering
\begin{minipage}{0.47\textwidth}
\setlength{\unitlength}{1cm}
\begin{picture}(0,0)
    \put(-0.36,6.3){\small\textbf{(a)}}
\end{picture}
\includegraphics[width=\linewidth]{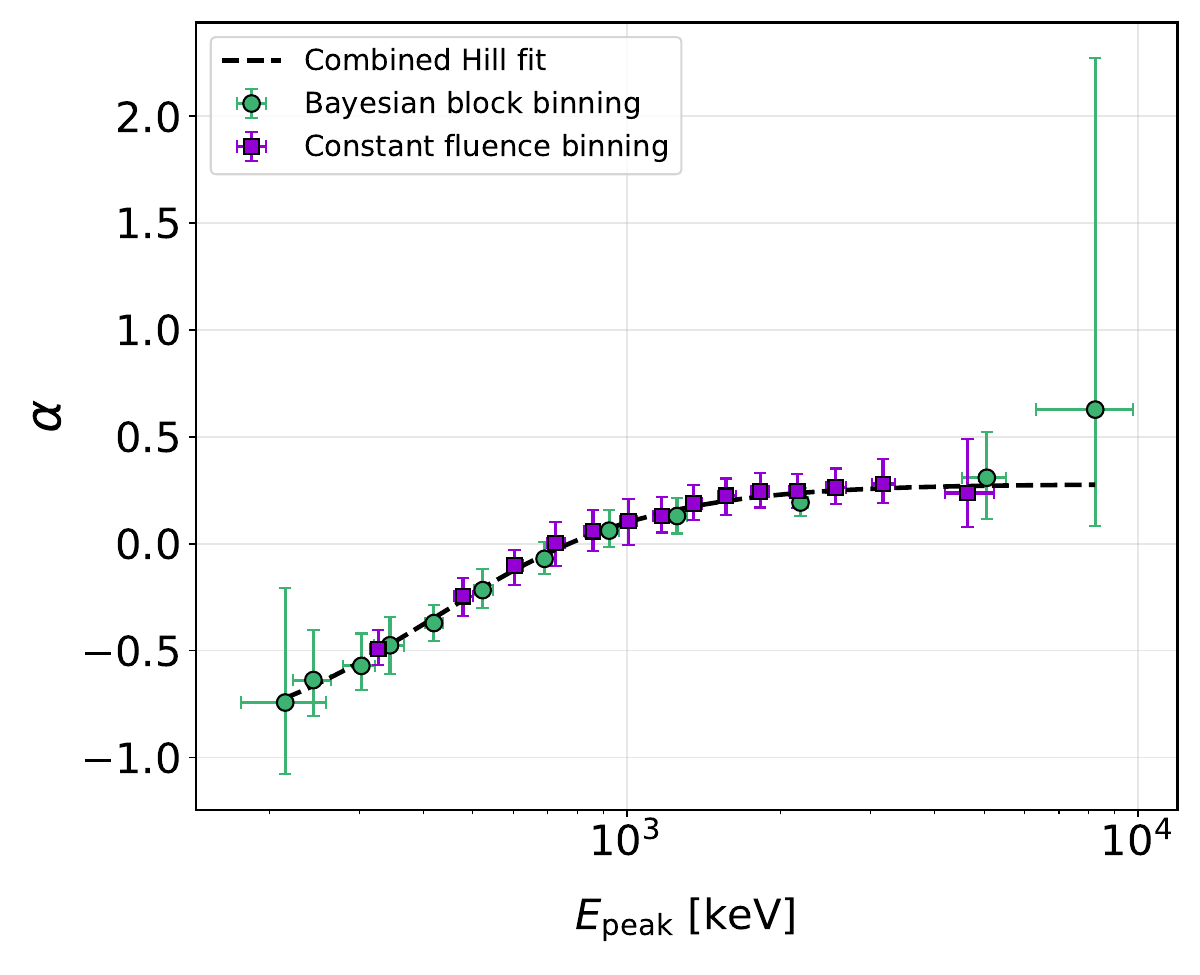}
\end{minipage}
\hspace{0.04\textwidth}
\begin{minipage}{0.47\textwidth}
\setlength{\unitlength}{1cm}
\begin{picture}(0,0)
    \put(-0.38,6.3){\small\textbf{(b)}}
\end{picture}
\includegraphics[width=\linewidth]{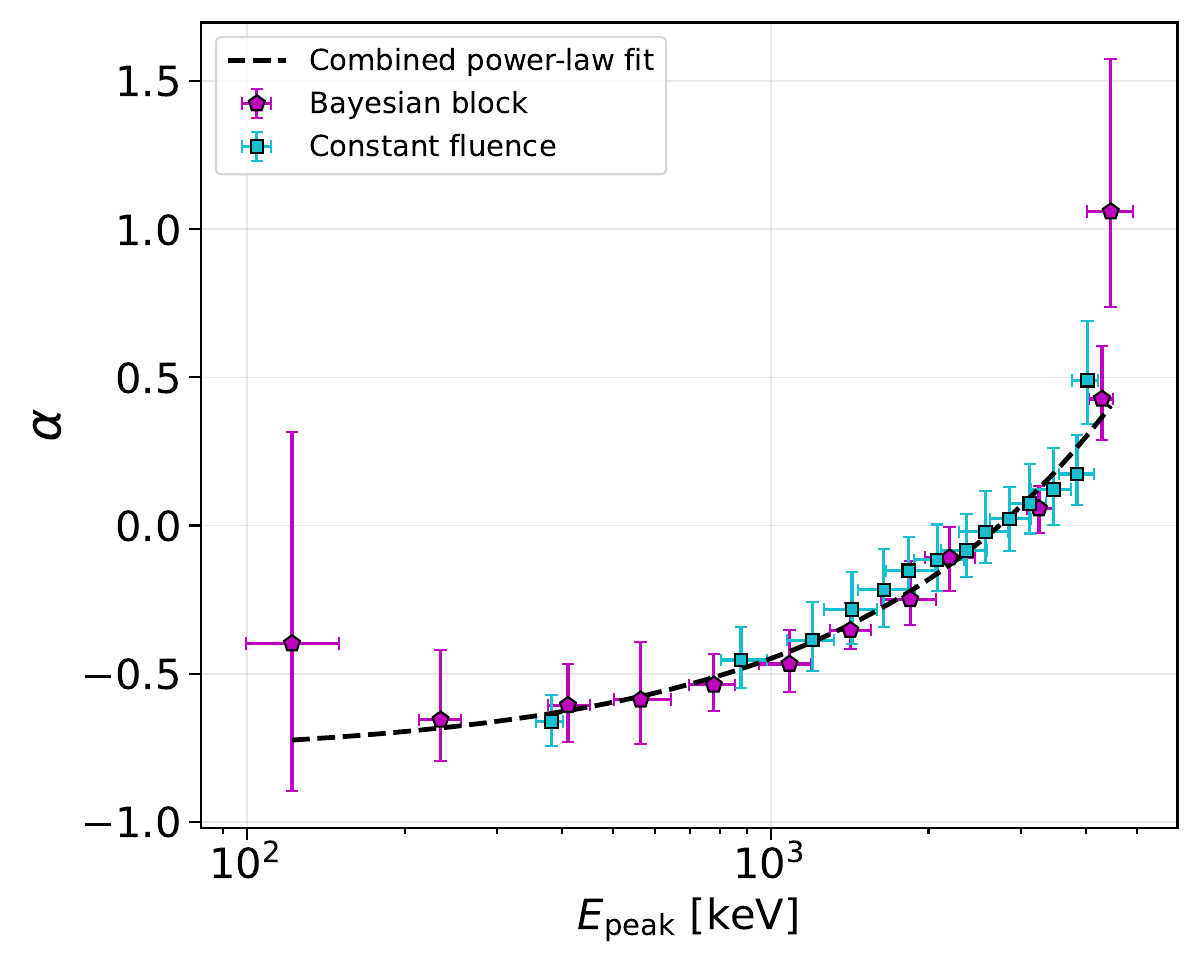}
\end{minipage}
\caption{(a) $E_{\rm peak}$--$\alpha$ correlation for the hard-to-soft $E_{\rm peak}$ evolution case. Green circles and violet squares represent the results obtained using Bayesian block and constant-fluence binning, respectively.
(b) Same as panel (a), but for the intensity-tracking $E_{\rm peak}$ evolution case. Magenta pentagons and cyan squares denote the Bayesian block and constant-fluence binning results, respectively.}
\label{Corr_HTS}
\end{figure}
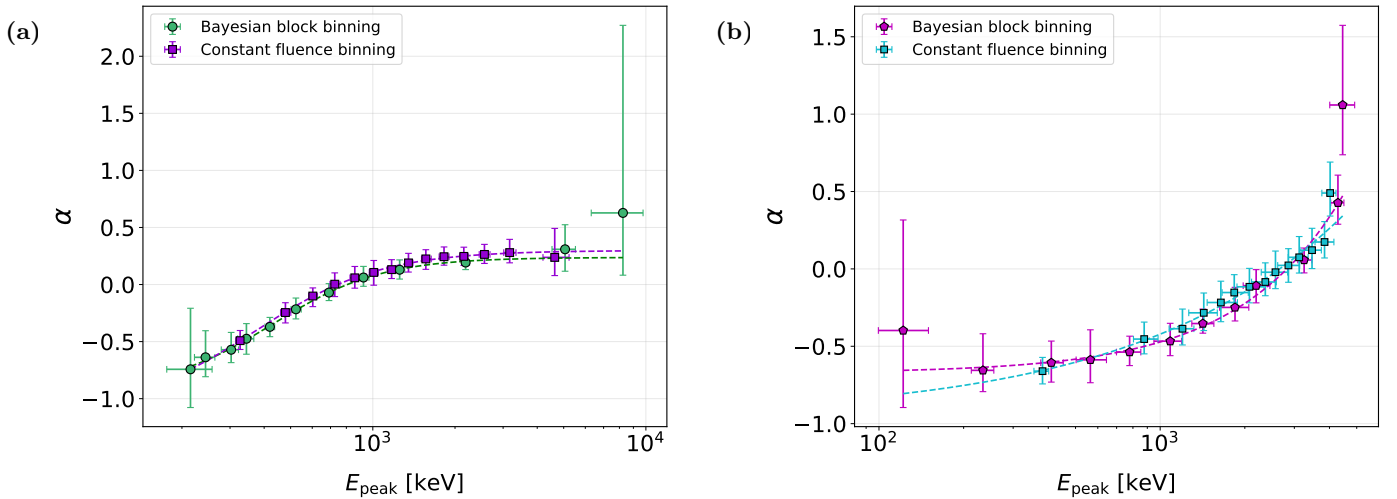

After applying the two binning techniques to each of the $E_{\rm peak}$ evolution patterns, we obtain four distinct cases in total. For each 
case, the values of $E_{\rm peak}$ and $\alpha$ derived from the Band function fits described in section \ref{spec_model_band}, are used to investigate the $E_{\rm peak}$–$\alpha$ correlation. We first visualize this relation using scatter plots for 
all four cases, as shown in Figures \ref{Corr_HTS}. In all cases, the plots reveal a clear positive correlation, irrespective of the assumed $E_{\rm peak}$ evolution pattern and the binning method. To quantitatively assess the strength of this correlation, we further compute the Pearson correlation coefficient and the Spearman rank correlation coefficient ($\rho$) which respectively measure the linear association and degree of monotonicity between $E_{\rm peak}$ and $\alpha$. The results are summarized in the Table \ref{tab:corr_summary}.\\
The Pearson correlation coefficient is $\geq 0.7$ and the Spearman rank coefficient is $> 0.8$ in all four cases, indicating a strong and monotonic positive correlation between $E_{\rm peak}$ and $\alpha$. \\ 

The $E_{\rm peak}$--$\alpha$ relation (Figure \ref{Corr_HTS}) exhibits distinct behaviours for the two $E_{\rm peak}$ evolution scenarios: the hard-to-soft case shows a power-law like evolution that gradually approaches an asymptotic 
$\alpha$ value, whereas the intensity-tracking case primarily exhibits a power-law behaviour without showing a clear transition toward saturation. To capture these different behaviours within a single generic functional form, we fit the $E_{\rm peak}$--$\alpha$ scatter points with a generalized Hill-type function, as shown by the dashed curves in Figures~\ref{Corr_HTS}(a) and \ref{Corr_HTS}(b). We have combined the data points obtained from both the binning methods for fitting in each of the $E_{\rm peak}$ cases. The mathematical expression of the Hill-type function is given below :

\begin{equation}
\alpha(E) = \alpha_{\min} + 
\frac{\alpha_{\max} - \alpha_{\min}}
{1 + \left(\frac{E_b}{E}\right)^k}
\label{Hill_func}
\end{equation}

where $\alpha_{\min}$ and $\alpha_{\max}$ represent the lower and upper asymptotic limits of the spectral index $\alpha$, respectively. The parameter $E_b$ denotes the characteristic transition energy that marks the change between the two asymptotic regimes, while $k$ is the curvature (or steepness) parameter that determines how rapidly the transition occurs between $\alpha_{\min}$ and $\alpha_{\max}$. 
\\The fit parameter values for all the 4 cases are summarized in the Table \ref{tab:hill_fit_params}. Unlike the hard-to-soft evolution case, the $E_{\rm peak}$--$\alpha$ scatter plot for the intensity-tracking case displays only a monotonic rising trend without a clear break or saturation. As a result, the Hill 
function parameters describing the saturation and the transition in slope $\alpha_{\rm max}$, $E_{b}$, and $k$ cannot be reliably constrained, as evident from the best-fit parameter values reported in Table~\ref{tab:hill_fit_params}. Over the range 
sampled by the data, the Hill function consequently reduces effectively to a single power-law relation. Thus, the large uncertainties in these parameters do not indicate a failure of the functional description; rather, they reflect the absence of a transition to the asymptotic regime, as a result the Hill function effectively reduces to a simple power-law relation.
For the hard-to-soft $E_{\rm peak}$ evolution case using Bayesian block binning (Case~2), the first main time bin contains relatively low energy flux and photon statistics, as evident from Figure~\ref{fig:Flux_Epeak_Temp_Binning}(b) and Figure~\ref{Simul_spectra_BA_m1_f12}. This bin also corresponds to the phase where $E_{\rm peak}$ is expected to attain its maximum value. As shown in Figure~\ref{Simul_spectra_BA_m1_f12} (Appendix~\ref{Extra_simul}), the simulated spectrum is intrinsically broad around the spectral peak, and additional spectral smearing further broadens the spectrum. We overplot the Band function using the median values of the fitted parameter distributions for the corresponding main bin. The comparison clearly shows that the curvature of the Band function is insufficient to reproduce the broad peak of the ICS simulated spectrum. As a result, the Band function parameter $E_{\rm peak}$ is not well constrained and tends to be overestimated. Consequently, the fitted value of $E_{\rm peak}$ is found to be $\sim 8000~\mathrm{keV}$, whereas the expected value is $\sim 4300~\mathrm{keV}$. In addition, the low photon statistics in this bin lead to large uncertainties in the fitted values of $\alpha$.

A similar effect is observed in the final main bins of Cases~2 and 4 (i.e., the Bayesian block binning cases), where low photon counts lead to large uncertainties in the estimated $\alpha$ values. Excluding these low-count intervals, both $E_{\rm peak}$ and $\alpha$ remain well constrained across all other time bins in the four cases summarized in Table~\ref{tab:corr_summary}.

Figure~\ref{Corr_HTS}(a) further reveals a noticeable change in the slope of the $E_{\rm peak}$--$\alpha$ correlation around $E_{\rm peak} \sim 1000$~keV, particularly in the hard-to-soft evolution scenario. The physical interpretation of this behaviour is discussed in Section~\ref{Slope_change}.

\begin{figure}[htbp]
\centering
\begin{minipage}{0.47\textwidth}
\setlength{\unitlength}{1cm}
\begin{picture}(0,0)
    \put(-0.36,6.3){\small\textbf{(a)}}
\end{picture}
\includegraphics[width=\linewidth]{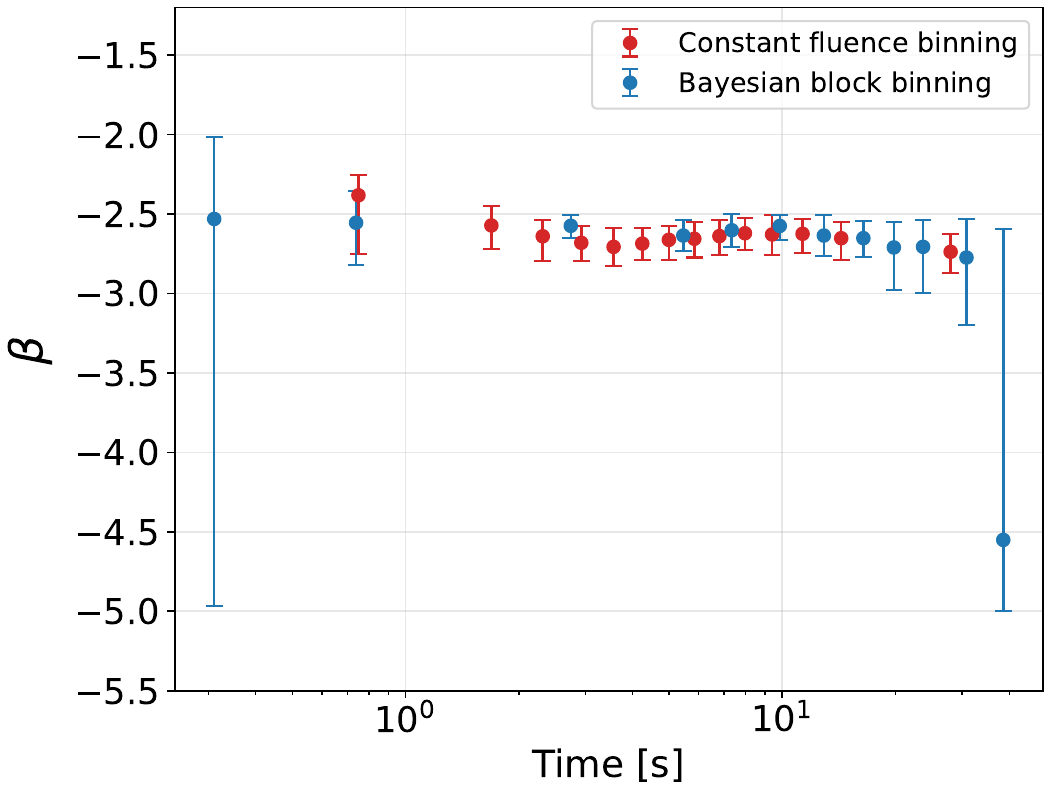}
\end{minipage}
\hspace{0.04\textwidth}
\begin{minipage}{0.47\textwidth}
\setlength{\unitlength}{1cm}
\begin{picture}(0,0)
    \put(-0.38,6.3){\small\textbf{(b)}}
\end{picture}
\includegraphics[width=\linewidth]{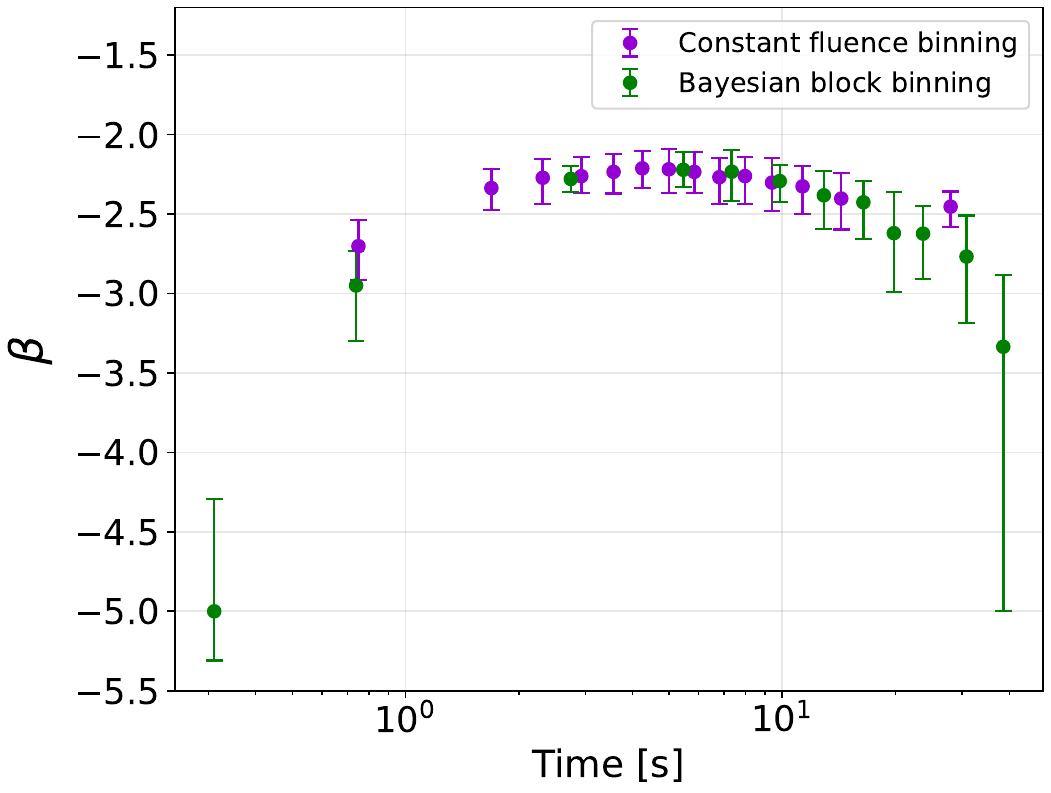}
\end{minipage}
\caption{Temporal evolution of the Band fit parameter, $\beta$, representing the high energy power law index (a) for the hard-to-soft $E_{\rm peak}$ evolution case (b) for the intensity tracking $E_{\rm peak}$ evolution case.}
\label{Beta_HTS_IT}
\end{figure}

As an additional consistency check of the adopted electron distribution, we examine the behaviour of the high-energy index $\beta$. In the present optically thin ICS model, the high-energy spectral shape is primarily determined by the power-law 
tail of the electron distribution. We adopt a constant electron power-law index, $\delta=2.25$, throughout the pulse and therefore expect little temporal evolution in $\beta$. The Band-function fits to our simulated spectra indeed yield nearly 
constant $\beta$ values, with average values of $\sim-2.6$ and $\sim-2.3$ for the hard-to-soft and intensity-tracking $E_{\rm peak}$ evolution cases, respectively, as shown in Figure~\ref{Beta_HTS_IT}. The first point in 
Figure~\ref{Beta_HTS_IT}(b) shows an unusually hard $\beta$, primarily due to limited photon statistics at high energies in that time bin, making the index poorly constrained. Excluding 
this point, $\beta$ remains approximately steady throughout the pulse. This behaviour is expected because, unlike $E_{\rm peak}$ and $\alpha$, the high-energy component remains within the observed energy range throughout the pulse and is governed primarily by the fixed electron power-law index.

The emphasis of the present study is on the $E_{\rm peak}$--$\alpha$ correlation, since these parameters are generally better constrained in time-resolved GRB spectral analysis owing 
to the higher photon statistics around the spectral peak and at lower energies. In contrast, $\beta$ is often poorly constrained because of the limited photon statistics at high energies. Furthermore, within the IC framework considered here, the high-
energy spectral shape is based on the assumed electron power-law index, while alternative electron distributions, such as a predominantly heated Maxwellian population expected in slow-
heating scenarios \citep{Ghisellini&Celotti1999,Pe'er2006a}, can lead to substantially different high-energy spectral behaviour like a cutoff. In addition, similar high-energy power-law spectra can also arise from other radiation processes, such as 
synchrotron emission, depending on the underlying electron power-law distribution \citep{Sari1998}. We, therefore, do not use $\beta$ as a primary diagnostic in the present study, but restrict its treatment to the consistency check presented above.

\begin{table}[htbp]
\centering
\caption{Correlation analysis between $E_{\text{peak}}$ and $\alpha$ for different evolution patterns and binning methods.}

\begin{tabular*}{\textwidth}{@{\extracolsep{\fill}}|c|c|c|c|c|}
\hline\hline
\textbf{Case} & \textbf{$E_{\text{peak}}$ Evolution} & \textbf{Binning Method} & \textbf{Pearson} & \textbf{Spearman ($\rho$)} \\
\hline\hline
1 & Hard-to-soft & Constant fluence & 0.70 (p = 5.4 $\times 10^{-3}$) & 0.96 (p = 7.4 $\times 10^{-11}$) \\
\hline
2 & Hard-to-soft & Bayesian blocks & 0.90 (p = 1.0 $\times 10^{-4}$) & 0.99 (p = 1.3 $\times 10^{-10}$) \\
\hline
3 & Intensity tracking & Constant fluence & 0.97 (p = 6.1 $\times 10^{-9}$) & 0.99 (p = 1.4 $\times 10^{-11}$) \\
\hline
4 & Intensity tracking & Bayesian blocks & 0.93 (p = 8.2 $\times 10^{-6}$) & 0.89 (p = 8.3 $\times 10^{-5}$) \\
\hline
\end{tabular*}

\label{tab:corr_summary}
\end{table}

\section{Discussions}
\label{Discussion}
\subsection{Origin of the correlation}
\label{Corr_disc_1}
Across an individual emission pulse, the Band-function fits to the simulated IC spectra exhibit a clear positive correlation between $E_{\rm peak}$ and $\alpha$. When $E_{\rm peak}$ attains very high values, the fitted low-energy spectral index becomes extremely 
hard, with $\alpha > 0$ and in some cases approaching the Planck limit ($\alpha \sim +1$). As $E_{\rm peak}$ evolves toward softer energies (typically a few hundred keV), $\alpha$ correspondingly softens to values near $-0.65$.

This behaviour arises from the manner in which the evolving IC spectrum is represented by the Band function within the finite energy range of {\it Fermi} GBM. During the high-flux, high-$E_{\rm peak}$ phase, both the thermal seed-photon component and the Comptonized 
component lie well within the detector energy window. In this regime, the Band $E_{\rm peak}$ primarily traces the Comptonized spectral hump, while the low-energy power-law segment effectively models the thermal portion of the spectrum. Consequently, the fitted $\alpha$ values become very hard and may even approach the Planck limit.

As the pulse evolves, both the flux and $E_{\rm peak}$ decrease, shifting the entire IC spectrum toward lower energies and lower normalization. The thermal component then weakens and gradually approaches, or falls below, 
the lower bound of the {\it Fermi} GBM energy range. As a result, the observed spectrum becomes increasingly dominated by the Comptonized component. In this phase, the low-energy power-law segment of the Band function no 
longer traces the thermal component, but instead fits the low-energy curvature of the Comptonized spectrum itself, naturally producing softer $\alpha$ values at later times. Thus, the observed $\alpha$--$E_{\rm peak}$ evolution arises from the combined effects of (i) the physical IC spectral evolution, and  (ii) the manner in which empirical models such as the Band function approximate the underlying radiation physics within a finite detector bandpass.\\

In this study, we employed two binning methods, namely constant fluence binning and Bayesian block binning, to construct time-resolved bins. We find that the choice of binning method does not broadly affect the $E_{\rm 
peak}$--$\alpha$ correlation. Time bins (main bins) with broader intervals encompass a larger number of fine bins, which can lead to a softening of the $\alpha$ 
value due to spectral smearing within those main bins. However, this softening effect is expected to be significant primarily in the last bins where the flux is 
low. In practice, the impact of this effect in the final bins does not substantially influence the overall time-resolved $E_{\rm peak}$--$\alpha$ correlation.


\subsection{Can the $E_{\rm peak}$-$\alpha$ correlation together with the $\alpha$ parameter serve as a diagnostic for IC?}
\label{Diag_discussion}
The time-resolved spectral parameters $E_{\rm peak}$ and $\alpha$ are generally well constrained in observed GRB spectra, making their correlation a useful probe of the underlying radiation mechanism. Our simulations show 
that, within the optically thin inverse Compton scattering (ICS) scenario, $E_{\rm peak}$ and $\alpha$ exhibit a monotonic positive correlation throughout the pulse evolution. As $E_{\rm peak}$ evolves from hard to 
soft energies with time, $\alpha$ correspondingly evolves from very hard, Planck-like values to progressively softer values. During the later stages of the pulse, $\alpha$ can even become softer than $-0.67$, the so-called synchrotron line of death \citep{Preece1998}.

The observed positive $E_{\rm peak}$--$\alpha$ correlation can be understood as a consequence of the evolution of the spectral peak through the finite energy band of the {\it Fermi} detectors during an emission pulse. A similar effect has also been reported for BATSE-
detected GRBs \citep{Lloyd_Petrosian2000}. Therefore, the presence of a positive $E_{\rm peak}$--$\alpha$ correlation alone cannot be regarded as a unique signature of a particular radiation mechanism 
and may also arise in other emission scenarios exhibiting an evolving spectral peak. The more relevant question is whether the range of $\alpha$ values and their systematic evolution with 
$E_{\rm peak}$, in conjunction with the positive correlation, can provide a discriminator among different radiation mechanisms. We, therefore, examine below the expected spectral behaviour of several commonly considered GRB radiation scenarios.

In the standard optically thin synchrotron model, assuming a constant magnetic field and an isotropic pitch-angle distribution \citep{Sari1998,Burgess2014}, the low-energy spectral index cannot be harder than $\alpha=-0.67$, corresponding to the well-known 
synchrotron line of death. Consequently, even if the evolution of the spectral peak through the observing band produces a positive $E_{\rm peak}$--$\alpha$ correlation during a synchrotron pulse, the associated $\alpha$ values are expected to remain at or below this limit, i.e., $\alpha\leq -0.67$.

A modified synchrotron scenario involving a decaying magnetic field, from initially strong to progressively weaker fields, can produce a temporal transition from fast- to slow-cooling synchrotron emission. 
In this case, the spectrum may evolve from $\alpha\lesssim -1.5$ during the fast-cooling phase toward harder values approaching the synchrotron limit of $\alpha=-0.67$ as the magnetic field decreases \citep{Daigne_Bosnjak2025}. Thus, the expected $\alpha$ range remains 
predominantly below the synchrotron line of death. Similarly, in the marginally fast-cooling synchrotron scenario, where $\nu_m$ lies just above the cooling frequency, the low-energy spectral index remains constrained to $\alpha\leq -2/3$ \citep{Daigne2011,Beniamini_etal_2018}.

Anisotropic pitch-angle distributions provide another modification to the synchrotron scenario and can produce harder low-energy spectral indices, $\alpha>-0.67$ \citep{Lloyd_Petrosian2000,Goto_2022}. 
However, the hardest asymptotic low-energy index attainable in this scenario is approximately $\alpha\sim0$. Thus, while pitch-angle effects can account for values harder than the standard synchrotron 
limit, they do not naturally extend to the significantly harder values of $\alpha>0$ obtained in the optically thin IC scenario considered in this work and also reported in \citealt{Bordoloi_Iyyani_2025}.

Two-component photospheric models provide another possible route to harder observed values of $\alpha$. In such models, a thermal blackbody (BB) component associated with non-dissipative photospheric 
emission is combined with a non-thermal component, often attributed to optically thin synchrotron emission \citep{Guiriec2011,Guiriec_etal_2015}. When the resulting spectrum is fitted with a single Band function, the inferred $\alpha$ can become harder than the intrinsic value of the non-thermal component, 
depending on the relative contribution of the BB and non-thermal components. This scenario can therefore also produce $\alpha>-0.67$. A possible distinction can be obtained through explicit two-component modelling. If the non-thermal component is synchrotron emission, 
fitting the spectrum with Band+BB should recover a Band component with the softer spectral indices expected from synchrotron emission. In contrast, if the non-thermal component originates from IC scattering, the Band component would obtain harder values, including 
$\alpha>0$, even after the thermal contribution is explicitly accounted for by the BB component \citep{Iyyani_etal_2015,Sharma_etal_2020}.

Taken together, these comparisons indicate that the positive $E_{\rm peak}$--$\alpha$ correlation by itself is not unique to IC 
scattering. However, the combination of this correlation with the characteristic range and systematic evolution of $\alpha$, particularly the occurrence of $\alpha>0$ during the evolution of an emission pulse or episode, provides a potentially discriminating 
signature of the optically thin IC scenario. In our simulations, the IC spectrum naturally evolves from very hard low-energy slopes, including $\alpha>0$, at high $E_{\rm peak}$ toward progressively softer values as $E_{\rm peak}$ decreases. We therefore propose that 
the joint $E_{\rm peak}$--$\alpha$ evolution and the associated $\alpha$ range can be used as a diagnostic for assessing the viability of IC scattering as the underlying radiation mechanism.

The present simulations specifically consider optically thin ICS above the photosphere, for which the evolution of $E_{\rm peak}$ and $\alpha$ is determined by the Comptonisation of the photospheric seed photons. A similar low-energy spectral behaviour may also be expected if the dissipation of jet kinetic 
energy occurs localised below the photosphere and IC is the dominant radiation process, where the spectral peak and low-energy slope can retain the characteristic signatures of the Comptonised radiation \citep{Peer&Waxman2005,Iyyani_etal_2015}. However, in the sub-photospheric case, multiple Compton scatterings and 
radiative thermalisation till the photopshere can substantially modify the high-energy part of the spectrum, depending on the location and optical depth of the dissipation region. Thus, while the $E_{\rm peak}$--$\alpha$ evolution may provide a diagnostic of an IC-dominated origin, the high-energy spectral component can provide a discriminator between optically thin and sub-photospheric dissipation scenarios \citep{Bordoloi_Iyyani_2025,Ahlgren_etal_2015}.

\subsection{Observed Time-resolved $E_{\rm peak}$--$\alpha$ Correlations}

Observational studies such as \citet{Kaneko2006} have shown that nearly $26\%$ of BATSE GRBs exhibit a positive correlation between the time-resolved $E_{\rm peak}$ and $\alpha$ parameters within individual pulses. In our simulations, we similarly find a monotonic positive correlation between $E_{\rm peak}$ and $\alpha$, independent of the adopted binning method or the assumed temporal evolution of $E_{\rm peak}$.

To further investigate this behaviour in observed {\it Fermi} GBM GRBs, we analyzed the results of the time-resolved Band-function fits reported in \citet{Bordoloi_2026} for a sample of 41 GRBs. These bursts were selected based on the criteria that the time-integrated spectra satisfy a fluence $> 10^{-5}~{\rm erg\,cm^{-2}}$, low-energy index 
$+1 \ge \alpha > -0.5$, and high-energy index $-1.7 > \beta > -3.3$, while the peak-flux spectra satisfy $+1 \ge \alpha > -0.45$ and $-1.7 > \beta > -3.3$. Such spectral properties make these GRBs promising candidates for prompt emission dominated by inverse Compton scattering (ICS). The selected sample spans a 
wide range of temporal morphologies, including both single- and multi-pulse bursts. It is important to emphasize that this sample is not intended to be an unbiased representation of the overall GRB population. Rather, it is a targeted sample 
selected independently on the basis of spectral characteristics (hard low-energy spectral indices and evidence for a thermal component) that make these bursts potential candidates for 
prompt emission involving the ICS process \citep{Bordoloi_2026}. Consequently, this sample does not determine the occurrence rate of the $E_{\rm peak}$--$\alpha$ correlation among all GRBs detected by {\it Fermi}, but to examine whether the positive correlation predicted by the optically thin ICS model is preferentially observed in a sample where ICS is expected to play an important role as a potential radiation process.

\begin{figure*}[!ht]
    \centering
    \includegraphics[width=1.0\linewidth]{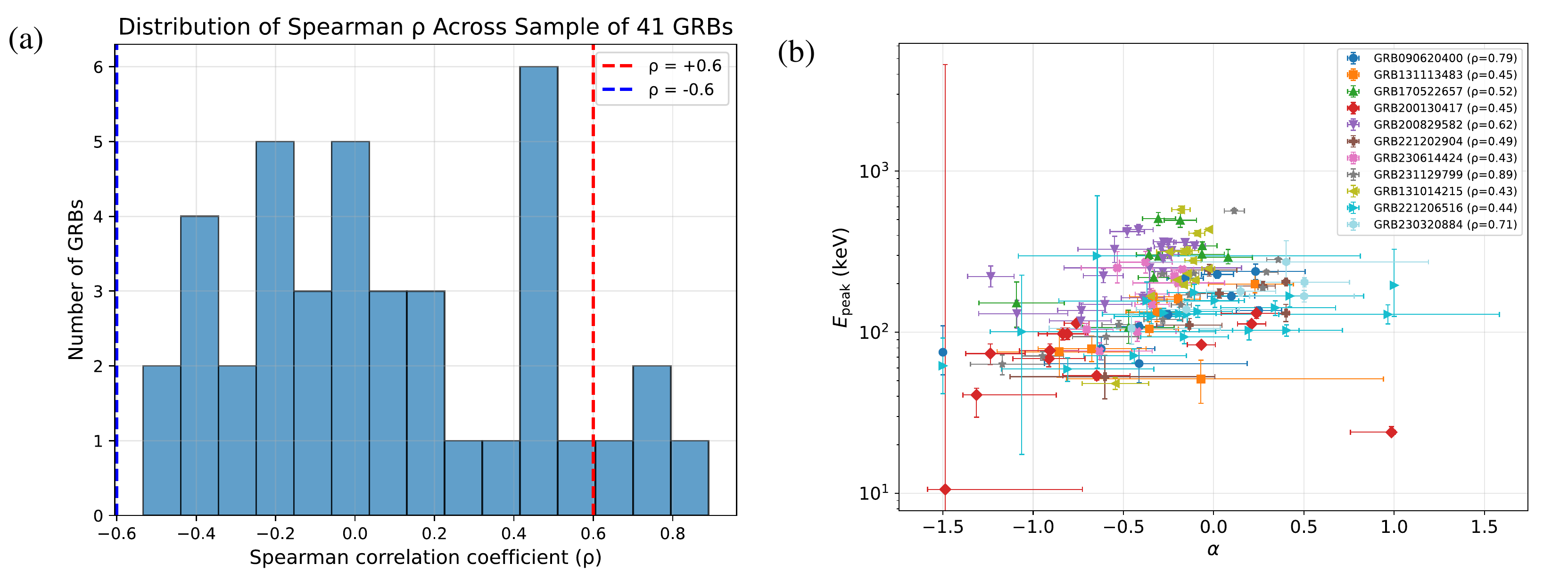}  
    \caption[$E_{peak}$- $\alpha$ correlation among the 41 GRB sample]{(a) The distribution of the Spearman correlation coefficients ($\rho$) for the $E_{\rm peak}$--$\alpha$ relation across a sample of 41 GRBs \citep{Bordoloi_etal_2026} is shown. The vertical red and blue dashed lines mark $\rho = +0.6$ and $\rho = -0.6$, respectively. (b) The corresponding $E_{\rm peak}$–$\alpha$ scatter plots for those GRBs exhibiting $\rho > +0.4$ indicating strong positive correlation are displayed.}
    \label{Epeak_alpha_correlations_sample}
\end{figure*}

For each GRB, we computed the Spearman rank correlation coefficient between the time-resolved $E_{\rm peak}$ and $\alpha$ values. The resulting distribution is shown in Figure~\ref{Epeak_alpha_correlations_sample}a. 
Approximately $34\%$ of the bursts exhibit a moderate to strong positive correlation ($\rho > +0.2$), with four GRBs showing particularly strong trends ($\rho > +0.6$). This fraction is broadly consistent with the results of \citet{Kaneko2006}. In contrast, nearly $29\%$ of the sample shows moderate negative correlations ($-0.6 < \rho < -0.2$), while the remaining $\sim 37\%$ exhibit no significant correlation ($-0.2 \leq \rho \leq +0.2$). 
Notably, none of the GRBs in the sample display a strong negative correlation ($\rho < -0.6$). The $E_{\rm peak}$--$\alpha$ scatter plots for GRBs exhibiting relatively stronger positive correlations ($\rho > +0.4$) are presented in Figure~\ref{Epeak_alpha_correlations_sample}b.

The presence of positive time-resolved $E_{\rm peak}$--$\alpha$ correlations in a significant fraction of the observed GRB sample, together with the fact that many of these bursts exhibit hard low-energy indices with $\alpha > -0.4$ and in some cases even $\alpha > 0$ (Figure~\ref{Epeak_alpha_correlations_sample}b), followed by a gradual softening of $\alpha$ as $E_{\rm peak}$ decreases, is broadly consistent with the trends 
obtained from our bottom-up ICS simulations. Here, $\alpha \sim -0.4$ represents the expected low-energy spectral index for non-dissipative photospheric emission modeled using the Band function \citep{Acuner_etal_2019}. The observed evolution from hard ($>$ 0) to softer $\alpha$ values therefore supports ICS as a viable candidate for the dominant radiation mechanism responsible for GRB prompt emission.

\subsection{Comparison with two composition hybrid jet models}
The hard $\alpha$ values are typically associated with a photospheric thermal component, while softer $\alpha$ values are generally interpreted as originating from non-thermal synchrotron radiation emitted from the optically thin region of the jet. There are several bursts, such as GRB 090719A, GRB 081224, and GRB 100707A, in which 
$\alpha$ evolves from hard to soft values together with the hard-to-soft evolution of $E_{\rm peak}$ within the same emission episode, as reported by \citealt{Li_2019}. This behavior indicates a positive correlation between 
the time-resolved $E_{\rm peak}$ and $\alpha$. In these bursts, the $\alpha$ values initially remain positive or well above the synchrotron line of death ($\alpha > -0.67$), and in the later phase decrease to values below this limit.\\
In the second part of their study, \citealt{Li_2020} interpreted this spectral evolution using a two-component hybrid jet model. In this scenario, the early phase of the prompt emission is dominated by a hot fireball photospheric component, while the later phase 
is dominated by a non-thermal component associated with a cold Poynting-flux-dominated jet. This interpretation implies that the GRB outflow contains both a thermal fireball component and a cold magnetically dominated component throughout the burst evolution, with the latter becoming increasingly dominant at later times.

As discussed in the previous subsection (\ref{Diag_discussion}), our simulations show that the observed hard-to-soft evolution of both $\alpha$ and 
$E_{\rm peak}$ can also arise naturally within the inverse Compton scattering (ICS) scenario. In contrast to the hybrid jet interpretation proposed by \citealt{Li_2020}, our results demonstrate that such spectral evolution can be reproduced within a single-
component baryonic fireball jet through ICS-dominated emission alone. This suggests that ICS within a baryon-dominated fireball outflow provides an alternative radiative framework capable of explaining the observed 
evolution of $\alpha$ and the associated behaviour of $E_{\rm peak}$.

\subsection{Effect of $E_{\rm peak}$ evolution on the  correlation}
\label{Slope_change}
Figure~\ref{Corr_HTS}(a) shows that the $E_{\rm peak}$--$\alpha$ correlation for Cases~1 and 2 exhibits a noticeable change in slope around $E_{\rm peak} \sim 1000$~keV. In contrast, no comparable feature is observed for Cases~3 and 4, shown in Figure~\ref{Corr_HTS}(b). Since this behaviour appears exclusively in the hard-to-soft $E_{\rm peak}$ evolution cases, independent of the adopted binning method, it suggests that the break originates from the nature of the imposed $E_{\rm peak}$ evolution itself.

As shown in Figure~\ref{fig:Flux_Epeak_Temp_Binning}(a) and (b), the hard-to-soft $E_{\rm peak}$ evolution is significantly steeper during the initial phase of the pulse (up to $\sim 6$~s, corresponding to $E_{\rm peak} \gtrsim 1000$~keV) than during the later phase. Consequently, the separation between consecutive time-resolved $E_{\rm peak}$ values is larger at early times and becomes progressively smaller as the evolution flattens. This transition in the temporal evolution of $E_{\rm peak}$ naturally produces the apparent change in slope observed in the $E_{\rm peak}$--$\alpha$ scatter plots of Figure~\ref{Corr_HTS}(a).

In contrast, for the intensity-tracking scenarios shown in Figure~\ref{fig:Flux_Epeak_Temp_Binning}(c) and (d), the temporal evolution of $E_{\rm peak}$ remains comparatively smooth throughout the pulse duration, without a pronounced change in slope. As a result, the corresponding $E_{\rm peak}$--$\alpha$ correlations do not exhibit any visible break structure.

These results indicate that the detailed temporal evolution of $E_{\rm peak}$ can influence the morphology of the time-resolved $E_{\rm peak}$--$\alpha$ correlation, particularly when $\alpha$ evolves gradually from hard to soft values. However, the overall positive nature of the correlation remains unchanged.

\subsection{Effect on $E_{\rm peak}$--$\alpha$ Correlation under Alternative $\Gamma$, $R_0$ and Pulse Evolution Scenarios}
\label{Deviation}
In our simulations, the temporal evolution of the GRB flux, $E_{\rm peak}$, bulk Lorentz factor ($\Gamma$), and nozzle radius ($R_0$) constitute the fundamental input parameters. The photospheric radius ($R_{\rm ph}$), comoving photospheric temperature ($kT_{\rm ph}^{\rm co}$), electron number density ($n_0$), and other physical quantities are then derived self-consistently from these inputs within the classical fireball framework. Below, we discuss alternative prescriptions for the evolution of $\Gamma$, $R_0$, and the pulse profile, and examine their implications for the resulting $E_{\rm peak}$--$\alpha$ correlation.

In the current framework, the bulk Lorentz factor $\Gamma$ decreases monotonically with time, while the nozzle radius $R_0$ evolves from $10^{6}$~cm to $10^{9}$~cm 
(Section~\ref{Nozzle_sec}). The combined evolution of these quantities determines the temporal evolution of the comoving seed-photon temperature, $kT_{\rm ph}^{\rm co}$, shown in Figure~\ref{phys_params_1}(e). This treatment differs from studies such as \citealt{Yan_2024}, where both $\Gamma$ and $R_0$ are assumed to remain constant throughout an individual GRB pulse. In this section, we examine the implications of alternate assumptions of $\Gamma$ and $R_0$ evolutions within our framework.

(i) Substituting constant values of $\Gamma$ and $R_0$ into Equations~(\ref{Photo_radius_1}) and (\ref{comoving_Trph}) gives
$kT_{\rm ph}^{\rm co} \propto L^{-5/12}$.
Thus, as the luminosity and $E_{\rm peak}$ decrease during the pulse, the comoving seed-photon temperature increases with time. This is contrary to the observed temporal evolution of the thermal component in GRBs, for which the observed temperature, $\Gamma kT_{\rm ph}^{\rm co}$, generally decreases with time \citep{Bordoloi_Iyyani_2025,Li_2020,Ryde2004}.

Furthermore, for fixed $\Gamma$, the increase in $T_{\rm ph}^{\rm co}$ during the pulse decay, together with the observed hard-to-soft evolution of $E_{\rm peak}$, requires the characteristic electron 
Lorentz factor $\gamma_e$ to decrease with time, since $E_{\rm peak}\propto\Gamma\gamma_e^2 kT_{\rm ph}^{\rm co}$. In the optically thin IC scenario considered here, previous studies \citealt{Bordoloi_Iyyani_2025,Bordoloi_2026} have found the 
scattering electrons to form a mildly relativistic heated Maxwellian population with $\gamma_e$ of order a few. A systematic decrease in $\gamma_e$ is therefore not naturally expected within this scenario. 
Furthermore, such an evolution would tend to produce a soft-to-hard evolution of $\alpha$, potentially resulting in a negative $E_{\rm peak}$--$\alpha$ correlation, contrary to the positive correlation commonly observed in GRB pulses.

\begin{figure*}[!ht]
    \centering
    \includegraphics[width=0.5\linewidth]{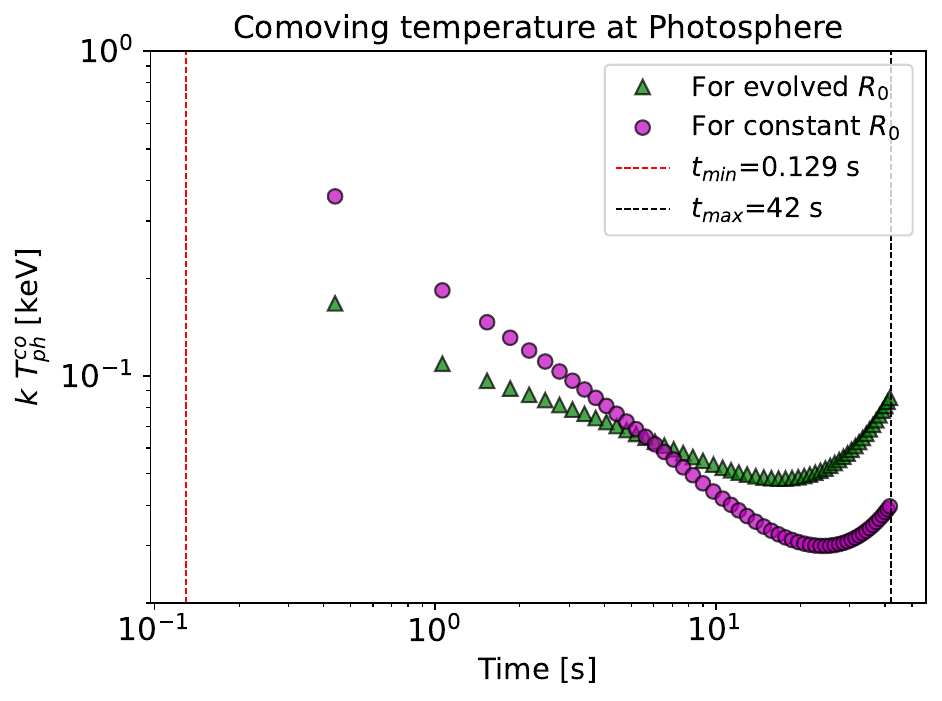}  
    \caption{Temporal evolution of the comoving photospheric temperature $kT_{\rm ph}^{\rm co}$ for case (a) where both $\Gamma$ and $R_0$ evolves, represented by the green triangles , case (b) where $R_0$ remains constant and $\Gamma$ evolves, represented by the magenta circles.}
    \label{Photo_temp_evolution}
\end{figure*}

(ii) We also examined a physically motivated alternative in which $R_0$ is kept constant while allowing $\Gamma$ to evolve. This assumption is motivated by studies of hybrid GRB outflows \citealt{Gao_etal_2015}, where a constant nozzle radius is adopted while the bulk Lorentz factor evolves with time according to the jet dynamics. In this case, the evolution of the seed-photon temperature, $kT_{\rm ph}^{\rm co}$, and consequently the electron thermal energy ($E_{\rm th}$), changes only marginally compared with our fiducial model employing an evolving $R_0$. As illustrated in Figure \ref{Photo_temp_evolution}, the temporal evolution of $kT_{\rm ph}^{\rm co}$ is nearly similar in both cases. Consequently, the seed-photon spectrum entering the ICS process remains largely unchanged, and we expect only a negligible impact on the simulated $\alpha$ evolution and the resulting $E_{\rm peak}$--$\alpha$ correlation for the same prescribed $E_{\rm peak}$ evolution.

These comparisons suggest that the temporal evolution of $\Gamma$ plays a more important role than that of $R_0$ in determining the evolution of the seed-photon temperature within our model. Moreover, the physically motivated evolution of both $\Gamma$ and $R_0$, expected in a baryonic fireball scenario where the central engine launches multiple shells with evolving outflow properties, naturally produces a decreasing seed-photon temperature during the pulse evolution. 

(iii) The adopted Norris pulse profile represents the commonly observed fast-rise, exponential-decay (FRED)-like temporal morphology of GRB pulses. A more symmetric pulse or one with a longer rise phase would primarily modify the relative duration of the rising and decaying phases and, consequently, the number of time-resolved bins sampled by the different 
binning schemes. However, this change in temporal sampling is not expected to alter the underlying spectral evolution inferred from the present study. In the present simulations, despite the relatively small number of bins in the rising phase, these bins do not exhibit a qualitatively different 
$E_{\rm peak}$-$\alpha$ evolution from that observed during the decay phase. The current study suggests that the inferred correlation is not significantly driven by the specific sampling of the rising phase. Therefore, we do not expect the correlation to be strongly dependent on the adopted Norris pulse profile. 

\section{Summary}
\label{conclusion}
In this work, we investigated the time-resolved $E_{\rm peak}$--$\alpha$ correlation using synthetic GRB spectra generated within an optically thin inverse Compton scattering (ICS)-dominated prompt emission framework. Physically motivated temporal evolutions of $E_{\rm peak}$, including both hard-to-soft and intensity-tracking 
behaviours, were implemented for an individual GRB pulse, and the underlying spectra were generated using the IC radiation framework. The simulated spectra were subsequently fitted with the empirical 
Band function in order to closely mimic standard observational analyses and extract the corresponding spectral parameters.

To examine the robustness of the inferred correlations against temporal binning effects, two commonly used time-resolved binning techniques were employed: constant-fluence binning, which 
preferentially samples the brighter high-flux phases of the burst, and Bayesian Block binning, which more uniformly captures both bright and faint intervals. Despite their distinct temporal 
sampling characteristics, both methods consistently produced a strong positive correlation between $E_{\rm peak}$ and $\alpha$.

The simulated spectra exhibit a systematic spectral evolution in which $\alpha$ evolves from  hard values ($\alpha >  0 $) at higher $E_{\rm peak}$ to progressively softer values approaching $\alpha \sim -0.65$ as $E_{\rm peak}$ decreases. This behaviour 
arises naturally from the evolution of the IC spectrum within the finite energy band of {\it Fermi} GBM. During the early high-flux phase, both the thermal seed-photon component and the Comptonized spectral hump lie within the detector energy range, causing the low-
energy Band index to effectively trace the thermal component and thereby attain very hard values. As the pulse evolves and the spectrum shifts toward lower energies and lower normalization, the 
thermal component gradually weakens and eventually moves close to, or below, the detector bandpass. The Band function then predominantly fits the low-energy curvature of the Comptonized component, naturally producing softer $\alpha$ values at later times.

Our results therefore show that the observed $E_{\rm peak}$--$\alpha$ evolution can arise from the combined effects of intrinsic IC spectral evolution and the manner in which the 
empirical Band function maps the underlying physical spectrum within a finite detector energy range. Overall, the study demonstrates that an ICS-dominated emission scenario naturally 
reproduces the positive pulse-level $E_{\rm peak}$--$\alpha$ correlation observed in many GRBs. In particular, the systematic evolution of $\alpha$ from hard, Planck-like values to softer 
slopes as $E_{\rm peak}$ decreases may serve as a useful diagnostic signature of ICS-dominated prompt emission. These results further suggest that such spectral evolution can be explained within a single-component baryonic fireball framework, providing an alternative to the two-component hybrid jet models often invoked to 
interpret this type of observed $\alpha$ evolution in GRBs.

\appendix 

\section{Table of the best fit parameters of the Hill function}
The Hill function fit parameter values for all 4 cases are reported in the Table \ref{tab:hill_fit_params}. The large uncertainties or unconstrained values of $\alpha_{\rm max}$, $E_b$, and $k$ in the intensity tracking case reflect the 
absence of a sampled transition toward the asymptotic regime; consequently, the Hill function effectively reduces to a simple power-law relation over the observed range (ref Section \ref{results} for more details). 
\begin{table*}
\centering
\caption{Best--fit parameters of the generalized Hill function describing the $E_{\rm peak}$--$\alpha$ correlation for different $E_{\rm peak}$ evolution patterns and binning methods.}
\begin{tabular}{|c|c|c|c|c|c|}
\hline\hline
\textbf{Case} & \textbf{$E_{\rm peak}$ Evolution} & \textbf{$\alpha_{\rm min}$} & \textbf{$\alpha_{\rm max}$} & \textbf{$E_b$ (keV)} & \textbf{$k$} \\
\hline\hline
1 & Hard-to-soft  
& $-0.98 \pm 0.60$ 
& $0.28 \pm 0.07$ 
& $416.50 \pm 195.30$ 
& $2.04 \pm  0.94$ \\
\hline
2 & Intensity tracking
& $-0.72 \pm 0.19$ 
& $2.00 \pm 5.85$ 
& $6226.77 \pm 18707.19$ 
& $1.22 \pm 1.12$ \\
\hline
\end{tabular}
\label{tab:hill_fit_params}
\end{table*}

\label{Input_parameter_intensity_tracking}
\section{Physical Parameter Space for ICS Simulation for Intensity-Tracking GRB Pulse}
\begin{figure}[htbp]
\centering

\begin{minipage}{0.45\textwidth}
\setlength{\unitlength}{1cm}
\begin{picture}(0,0)
    \put(-0.36,5.5){\small\textbf{(a)}}
\end{picture}
\includegraphics[width=\linewidth]{Evol_Gamma_CFB_HT.pdf}
\end{minipage}
\hspace{0.01\textwidth}
\begin{minipage}{0.45\textwidth}
\setlength{\unitlength}{1cm}
\begin{picture}(0,0)
    \put(-0.38,5.5){\small\textbf{(b)}}
\end{picture}
\includegraphics[width=\linewidth]{Evol_R0_CFB_HT.pdf}
\end{minipage}

\vspace{0.3cm}

\begin{minipage}{0.45\textwidth}
\setlength{\unitlength}{1cm}
\begin{picture}(0,0)
    \put(-0.36,5.5){\small\textbf{(c)}}
\end{picture}
\includegraphics[width=\linewidth]{Evol_luminosity_CFB_HT.pdf}
\end{minipage}
\hspace{0.01\textwidth}
\begin{minipage}{0.45\textwidth}
\setlength{\unitlength}{1cm}
\begin{picture}(0,0)
    \put(-0.38,5.5){\small\textbf{(d)}}
\end{picture}
\includegraphics[width=\linewidth]{Evol_n0_CFB_HT.pdf}
\end{minipage}

\vspace{0.3cm}
\begin{minipage}{0.45\textwidth}
\setlength{\unitlength}{1cm}
\begin{picture}(0,0)
    \put(-0.36,5.5){\small\textbf{(e)}}
\end{picture}
\includegraphics[width=\linewidth]{Evol_Tph_CFB_HT.pdf}
\end{minipage}
\hspace{0.01\textwidth}
\begin{minipage}{0.45\textwidth}
\setlength{\unitlength}{1cm}
\begin{picture}(0,0)
    \put(-0.38,5.5){\small\textbf{(f)}}
\end{picture}
\includegraphics[width=\linewidth]{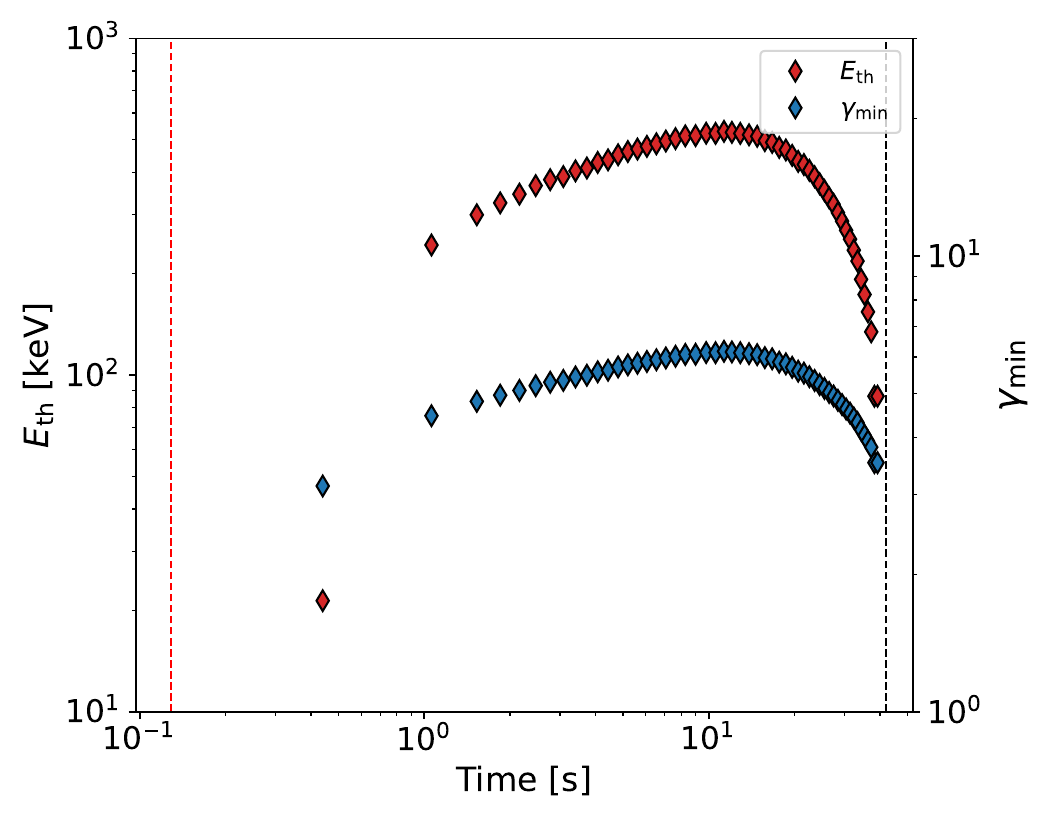}
\end{minipage}
\caption{The temporal evolution of the underlying physical parameters for the intensity tracking $E_{\rm peak}$ evolution. (a) bulk Lorentz factor, $\Gamma$. The magenta and blue dashed horizontal lines represent the maximum and minimum values of $\Gamma$; (b) Nozzle radius, $R_0$. The magenta and blue dashed horizontal lines represent the maximum and minimum values of $R_0$; (c) Luminosity, $L$; (d) Electron distribution normalisation, $n_0$, (e) Co-moving temperature at the photosphere, $kT_{\rm ph}^{\rm co}$ (in energy unit) and (f) electron thermal energy, $E_{th}$ (in red diamonds) and the minimum Lorentz factor of power law electrons, $\gamma_{min}$ (in blue diamonds) , are shown.}
\label{phys_params_3}

\end{figure}

\begin{figure}[htbp]
\centering
\begin{minipage}{0.47\textwidth}
\setlength{\unitlength}{1cm}
\begin{picture}(0,0)
    \put(-0.36,5.7){\small\textbf{(a)}}
\end{picture}
\includegraphics[width=\linewidth]{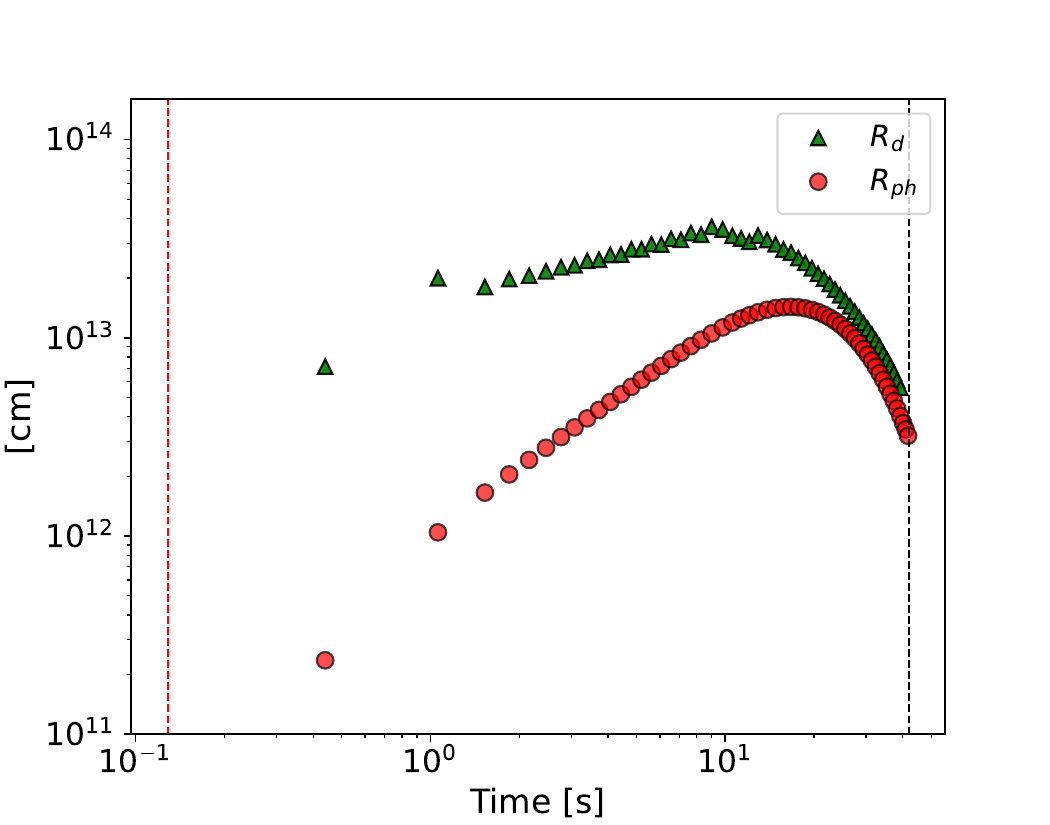}
\end{minipage}
\hspace{0.01\textwidth}
\begin{minipage}{0.47\textwidth}
\setlength{\unitlength}{1cm}
\begin{picture}(0,0)
    \put(-0.38,5.7){\small\textbf{(b)}}
\end{picture}
\includegraphics[width=\linewidth]{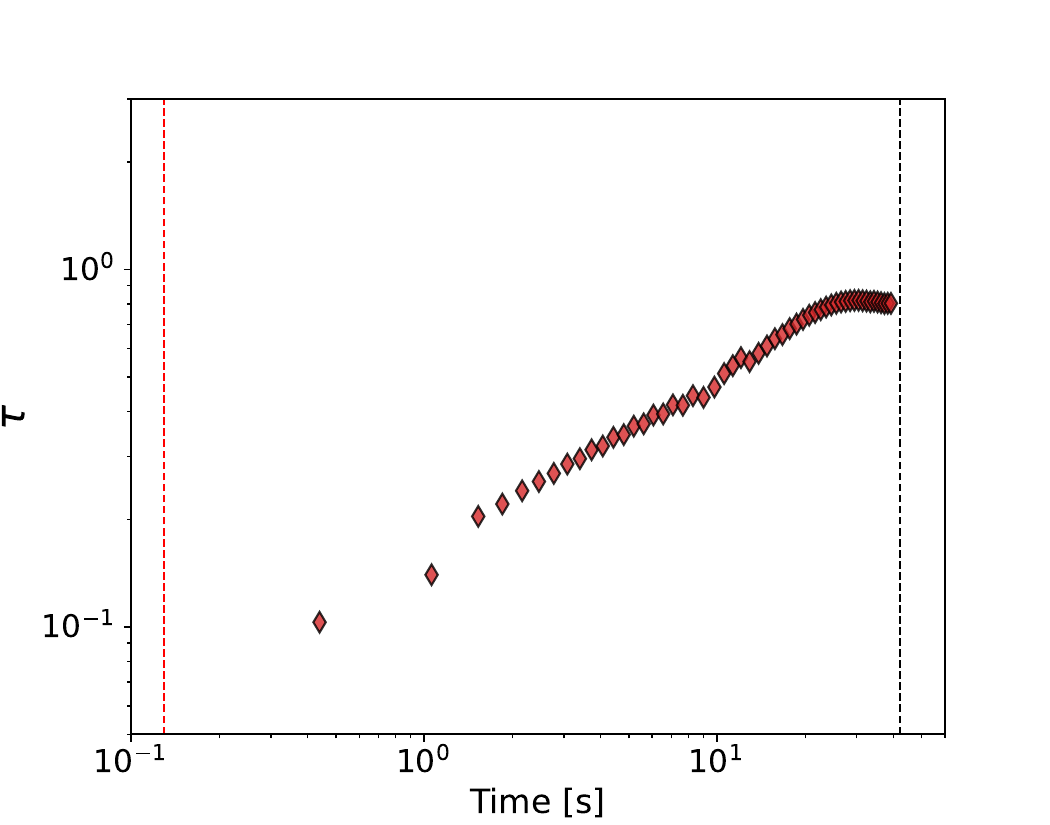}
\end{minipage}
\caption{The temporal evolution of the underlying physical parameters (a) dissipation radius, $R_d$ (green traingles) and photospheric radius $R_{ph}$ (red circles), and (b) optical depth, $\tau$ at the dissipation site in case of the intensity tracking $E_{peak}$ evolution scenario are shown.}
\label{phys_params_4}
\end{figure}

The underlying input parameter evolution adopted for simulating the ICS spectra in the intensity-tracking $E_{\rm peak}$ scenario is presented in Figures~\ref{phys_params_3} and \ref{phys_params_4}. This parameter space closely resembles that used for the hard-to-soft $E_{\rm peak}$ evolution case discussed in Section~\ref{Physical_para_setup}.  

\label{Input_parameter_intensity_tracking}


\section{Counts + Residual plots for case 2}
Case 2 corresponds to the scenario in which the $E_{\rm peak}$ evolution follows a hard-to-soft trend and the spectra are binned using the Bayesian block method. To evaluate the quality of the Band model fit, we present the counts-plus-residual plots for the peak bin of Case 2, allowing the fit quality to be assessed through visual inspection of the residuals.

\begin{figure}[htbp]
\centering


{\large\textbf{Hard-to-Soft $E_{\rm peak}$ evolution Case}}

\vspace{0.25cm}


\begin{minipage}{0.47\textwidth}
\centering
\setlength{\unitlength}{1cm}
\begin{picture}(0,0)
    \put(-4.7,-0.3){\small\textbf{(a)}}
\end{picture}
\includegraphics[width=\linewidth]{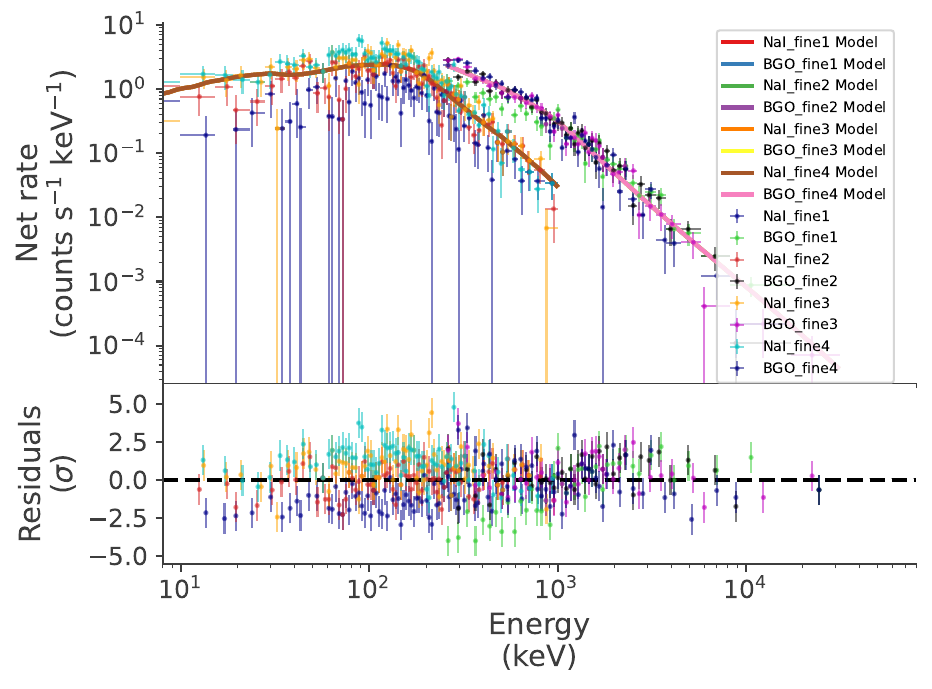}
\end{minipage}
\hspace{0.01\textwidth}
\begin{minipage}{0.42\textwidth}
\centering
\setlength{\unitlength}{1cm}
\begin{picture}(0,0)
    \put(-4.3,0.1){\small\textbf{(b)}}
\end{picture}
\includegraphics[width=\linewidth]{vFv_main_bin_3_HTS.pdf}
\end{minipage}


\vspace{0.6cm}


{\large\textbf{Intensity-Tracking $E_{\rm peak}$ evolution Case}}

\vspace{0.25cm}


\begin{minipage}{0.47\textwidth}
\centering
\setlength{\unitlength}{1cm}
\begin{picture}(0,0)
    \put(-4.7,-0.3){\small\textbf{(c)}}
\end{picture}
\includegraphics[width=\linewidth]{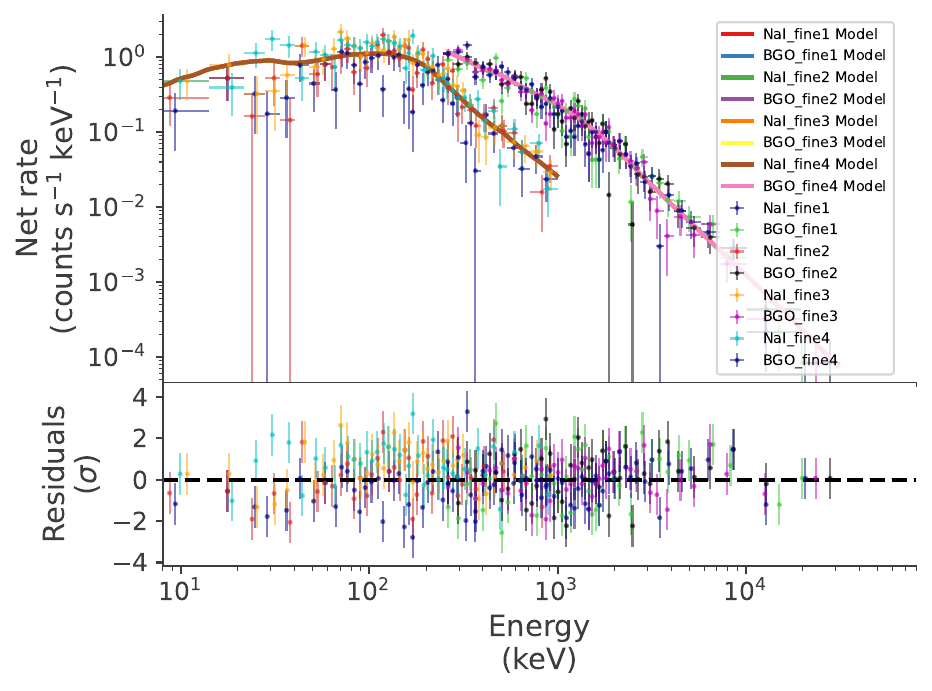}
\end{minipage}
\hspace{0.03\textwidth}
\begin{minipage}{0.42\textwidth}
\centering
\setlength{\unitlength}{1cm}
\begin{picture}(0,0)
    \put(-4.3,0.007){\small\textbf{(d)}}
\end{picture}
\includegraphics[width=\linewidth]{vFv_main_bin_3_IT_case.pdf}
\end{minipage}


\caption{
Counts-plus-residual plots (left panels) and corresponding $\nu F_{\nu}$ spectra (right panels) for the peak bin in the hard-to-soft and intensity-tracking spectral-evolution cases using Bayesian block binning.
}

\label{count_residual_plot}

\end{figure}

\section{Simulated ICS Spectra in the First Main Bin of the Norris Flux Profile (at $t = 0.3$ s) for Case 2}
\label{Extra_simul}

\begin{figure}[H]
    \centering
    \includegraphics[width=0.5\linewidth]{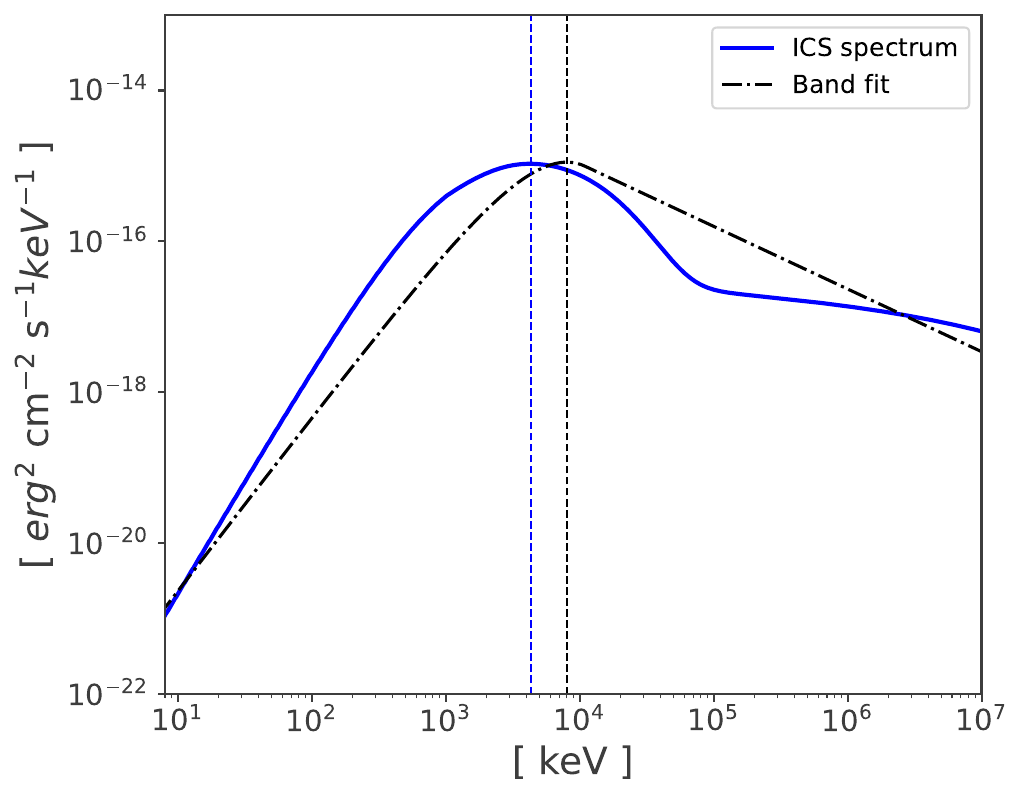} 
    \caption{The blue solid curve represents the ICS spectrum in the first main (broad) bin, simulated under a hard-to-soft $E_{\rm peak}$ evolution scenario using Bayesian block time binning, while the shaded dotted curve shows the corresponding best-fit Band function in the same bin. The blue and black dashed vertical lines represent the spectral peaks corresponding to the simulated ICS spectrum and the best-fit Band function, respectively. 
    }
    
    \label{Simul_spectra_BA_m1_f12}
\end{figure}

The time-resolved spectrum corresponding to the first time bin of the Norris pulse, exhibiting a hard-to-soft evolution of $E_{\rm peak}$ and obtained using the Bayesian block binning method, is shown by the blue curve in Figure~\ref{Simul_spectra_BA_m1_f12}. The black dash-dotted curve represents the corresponding best-fit Band function, overlaid on the ICS spectrum to illustrate that the Band function curvature is insufficient to adequately reproduce the broad peak of the ICS spectrum. The dashed vertical blue and black lines indicate the peak energies of the ICS spectrum and the fitted Band function, located at $\sim 4300$ keV and $\sim 8000$ keV, respectively.

\bibliography{Ref1}
\bibliographystyle{aasjournalv7}



\end{document}